\documentclass[11pt]{article}
\usepackage{appb}    %  ,epsfig
\usepackage{hyperref}

\input boxedeps
\SetepsfEPSFSpecial
\HideDisplacementBoxes

\def\slashii#1{\setbox0=\hbox{$#1$}             % set a box for #1
   \dimen0=\wd0                                 % and get its size
   \setbox1=\hbox{\sl/} \dimen1=\wd1            % get size of /
   \ifdim\dimen0>\dimen1                        % #1 is bigger
      \rlap{\hbox to \dimen0{\hfil\sl/\hfil}}   % so center / in box
      #1                                        % and print #1
   \else                                        % / is bigger
      \rlap{\hbox to \dimen1{\hfil$#1$\hfil}}   % so center #1
      \hbox{\sl/}                               % and print /
   \fi}                                         %
\def\slashiii#1{\setbox0=\hbox{$#1$}#1\hskip-\wd0\hbox to\wd0{\hss\sl/\/\hss}}
\def\vev#1{\langle #1\rangle_0}

\def\abs#1{\left| #1\right|}

\def\ltap{\mathop{\raisebox{-.4ex}{\rlap{$\sim$}} 
\raisebox{.4ex}{$<$}}}
\def\gtap{\mathop{\raisebox{-.4ex}{\rlap{$\sim$}} 
\raisebox{.4ex}{$>$}}}
\def\beq{\begin{equation}}
\def\eeq{\end{equation}}

\def\bea{\begin{eqnarray}}
\def\eea{\end{eqnarray}}
\def\bentarrow{\:\raisebox{1.3ex}{\rlap{$\vert$}}\!\rightarrow}

\def\dkp#1#2#3#4{
	\begin{equation}
	\begin{array}{r c l}
	#1 & \rightarrow & #2#3 \\
	 & & \phantom{\; #2}\bentarrow #4
	\end{array}
	\end{equation}
		}
\def\bothdk#1#2#3#4#5#6{
	\begin{equation}
	\begin{array}{r c l}
	#1 & \rightarrow & #2#3 \\
	 & & \:\raisebox{1.3ex}{\rlap{$\vert$}}\raisebox{-0.5ex}{$\vert$}%
	\phantom{#2}\!\bentarrow #4 \\
	 & & \bentarrow #5
	\end{array}\label{eq:#6}
	\end{equation}
		}
\def\twodk#1#2#3#4#5{
	\begin{equation}
	\begin{array}{l}
	#1\;#2 \\
	 \raisebox{1.3ex}{\rlap{$\vert$}}\raisebox{-0.5ex}{$\vert$}%
	\phantom{#2}\!\bentarrow #4\\
	\hspace{-2.2pt}\bentarrow #3 
	\end{array}\label{eq:#5}
	\end{equation}
		}

\headtitle{Higgs Physics}
\headauthor{C.~Quigg}

\def\url#1{\mbox{\href{#1}{\sf #1}}}
\def\urll#1#2{\mbox{\href{#1}{\sf #2}}}

\newcommand{\hepex}[1]{\mbox{\href{http://xxx.lanl.gov/abs/hep-ex/#1}{(hep-ex/#1)}}}
\newcommand{\hepph}[1]{\mbox{\href{http://xxx.lanl.gov/abs/hep-ph/#1}{(hep-ph/#1)}}}
\newcommand{\hepth}[1]{\mbox{\href{http://xxx.lanl.gov/abs/hep-th/#1}{(hep-th/#1)}}}
\newcommand{\heplat}[1]{\mbox{\href{http://xxx.lanl.gov/abs/hep-lat/#1}{(hep-lat/#1)}}}
\newcommand{\astro}[1]{\mbox{\href{http://xxx.lanl.gov/abs/astro-ph/#1}{(astro-ph/#1)}}}
\def\npbps#1#2#3{{\em Nucl. Phys. B (Proc. Supp.)\/} {\bf #1} (19#3) #2}
\def\prl#1#2#3{\frenchspacing{\it Phys. Rev. Lett. }{\bf #1}, #2 (19#3)}
\def\pr#1#2#3{\frenchspacing{\it Phys. Rev. D}{\bf #1}, #2 (19#3)}
\def\pl#1#2#3{\frenchspacing{\it Phys. Lett. }{\bf #1}, #2 (19#3)}
\def\np#1#2#3{\frenchspacing{\it Nucl. Phys. }{\bf #1}, #2 (19#3)}
\def\jmp#1#2#3{\frenchspacing{\it J. Math. Phys. }{\bf #1}, #2 (19#3)}
\def\app#1#2#3{\frenchspacing{\it Acta Phys. Polon. }{\bf #1}, #2 (19#3)}
\def\rmp#1#2#3{\frenchspacing{\it Rev. Mod. Phys. }{\bf #1}, #2 (19#3)}

\def\prep#1#2#3{\frenchspacing{\it Phys. Rep. }{\bf #1}, #2 (19#3)}
\def\arnps#1#2#3{\frenchspacing{\it Ann. Rev. Nucl. Part. Sci. }{\bf #1}, #2 (19#3)}
\def\ib#1#2#3{{\it ibid. }{\bf #1}, #2 (19#3)}

\def\sciam#1#2#3#4{{\it Sci. Am. }{\bf #1}, #2 (\ifcase#3\or January\or February\or March\or April\or May\or
June\or July\or August\or September\or October\or November\or December\fi, 19#4)}
\def\uspek#1#2#3#4#5#6{{\it Usp. Fiz. Nauk }{\bf #1}, #2 (19#3) [English translation: 
         {\it Sov. Phys.--Uspekhi }{\bf #4}, #5 (19#6)]}
\def\zp#1#2#3{{\it Z.~Phys. C}{\bf#1}, #2 (19#3)}

\def\ws{$SU(2)_L\otimes U(1)_Y$}
\def\sm{$SU(3)_c\otimes SU(2)_L\otimes U(1)_Y$}
\def\D{{\cal D}}
\def\L{{\cal L}}

\def\onetev{1-TeV scale}
\newcommand{\mumu}{$\mu^{+}\mu^{-}$ collider}

\newcommand{\etal}{{\em et al.}}
\newcommand{\ie}{{\em i.e.}}
\newcommand{\eg}{{\em e.g.}}
\newcommand{\cf}{{\em cf.\ }}
\newcommand{\gevcc}{\hbox{ GeV}\!/\!c^2}
\newcommand{\gev}{\hbox{ GeV}}

\newcommand{\evcc}{\hbox{ eV}\!/\!c^2}
\newcommand{\mev}{\hbox{ MeV}}

\newcommand{\mevcc}{\hbox{ MeV}\!/\!c^2}
\newcommand{\tev}{\hbox{ TeV}}
\newcommand{\tevcc}{\hbox{ TeV}\!/\!c^2}

\newcommand{\fb}{\hbox{ fb}}

\def\ltap{\mathop{\raisebox{-.4ex}{\rlap{$\sim$}} 
\raisebox{.4ex}{$<$}}}
\def\gtap{\mathop{\raisebox{-.4ex}{\rlap{$\sim$}} 
\raisebox{.4ex}{$>$}}}
\def\eqn#1{(\ref{#1})}

\newcommand{\cfrac}[2]{\textstyle \frac{#1}{#2}}

\begin{document}
    \preprint{FERMILAB-CONF-99/033-T}
\eqsec  % uncomment this line to get equations numbered by (sec.num)
\title{ELECTROWEAK SYMMETRY BREAKING \\ AND THE HIGGS SECTOR
\thanks{Presented at the XXVII International Meeting on Fundamental 
Physics, Sierra Nevada (Granada), Spain, 1 -- 5 February 1999.}}
\author{Chris QUIGG %\thanks{}
\address{
Theoretical Physics Department,\\ Fermi National Accelerator 
Laboratory, \\ P.O. Box 500, Batavia, Illinois 60510 USA \\
E-mail: \textsf{quigg@fnal.gov}
}}
\maketitle

\begin{abstract}
These three lectures review the state of our understanding of 
electroweak interactions and the search for the agent of electroweak 
symmetry breaking.  The themes of the lectures are \textit{(i)} the 
electroweak theory and its experimental status, \textit{(ii)} the 
standard-model Higgs boson, and \textit{(iii)} aspects of electroweak 
theory beyond the standard \ws\ model.
\end{abstract}
  
\section{Introduction}
The central challenge in particle physics is to explore the 1-TeV 
scale and elucidate the nature of electroweak symmetry breaking.  A 
key element in this quest is the search for the Higgs boson, the 
agent of electroweak symmetry breaking in the standard electroweak 
theory.  

Uncovering the secrets of the Higgs sector is the focus of much 
present and future experimental research.  The LEP~2 experiments are 
searching now for a light Higgs boson and for low-scale 
supersymmetry.  At the Tevatron, CDF and D\O\ will begin next year a 
high-luminosity run with considerable sensitivity to new physics, and 
offer promise for decisive light-Higgs searches in the future.  The 
Large Hadron Collider at CERN will bring extensive explorations of 
TeV-scale physics beginning in about 2005.  Linear colliders now on 
the drawing boards would offer complementary possibilities for the 
study of electroweak symmetry breaking.

These three lectures offer a short course in the current state of 
electroweak symmetry breaking and the Higgs sector.  The subject is 
vast, so many important topics will receive only a schematic 
treatment.  Complementary views of the electroweak panorama are to be 
found in other recent lecture notes 
\cite{Veltman:1997nm,Fayet:1998wb,Quigg:1996ew,Ellis:1998eh,spirz,Dawson:1998yi,okunfest}.
\section{The Electroweak Theory}
\subsection{Brief R\'esum\'e and Perspective {\protect{\label{sub:cache}}}}
Let us review the essential elements of the \ws\ electroweak 
theory \cite{GT,Herrero:1998eq,Gaillard:1998ui}.
The electroweak theory takes three crucial clues from experiment:
\begin{itemize}
    \item  The existence of left-handed weak-isospin doublets,
    \begin{displaymath}
        \left( 
                \begin{array}{c}
            \nu_{e}  \\
            e
        \end{array}
	\right)_{L} \qquad
	        \left( 
                \begin{array}{c}
            \nu_{\mu}  \\
            \mu
        \end{array}
	\right)_{L} \qquad
        \left( 
                \begin{array}{c}
            \nu_{\tau}  \\
            \tau
        \end{array}
	\right)_{L}
    \end{displaymath}
    and
    \begin{displaymath}
        \left( 
                \begin{array}{c}
            u  \\
            d^{\prime}
        \end{array}
	\right)_{L} \qquad
	        \left( 
                \begin{array}{c}
            c  \\
            s^{\prime}
        \end{array}
	\right)_{L} \qquad
        \left( 
                \begin{array}{c}
            t  \\
            b^{\prime}
        \end{array}
	\right)_{L}\; ; 
    \end{displaymath}

    \item  The universal strength of the weak interactions;

    \item  The idealization that neutrinos are massless.
\end{itemize}

To save writing, we shall construct the electroweak theory as it 
applies to a single generation of leptons.  In this form, it is 
neither complete nor consistent: anomaly cancellation requires that a 
doublet of color-triplet quarks accompany each doublet of 
color-singlet leptons.  However, the needed generalizations are simple 
enough to make that we need not write them out.

To incorporate electromagnetism into a theory of the weak 
interactions, we add to the $SU(2)_{L}$ family symmetry suggested by 
the first two experimental clues a $U(1)_{Y}$ weak-hypercharge phase 
symmetry.  We begin by specifying the fermions: a left-handed weak 
isospin doublet
\begin{equation}
{{\sf L}} = \left(\begin{array}{c} \nu_e \\ e \end{array}\right)_L
\end{equation}
with weak hypercharge $Y_L=-1$, and a right-handed weak isospin singlet
\begin{equation}
      {{\sf R}}\equiv e_R
\end{equation}
with weak hypercharge $Y_R=-2$.

The electroweak gauge group, \ws, implies two sets of gauge fields:
a weak isovector $\vec{b}_\mu$, with coupling constant $g$, and a
weak isoscalar
${{\mathcal A}}_\mu$, with coupling constant $g^\prime$. Corresponding
to these gauge fields are the field-strength tensors 
\begin{equation}
    F^{\ell}_{\mu\nu} = \partial_{\nu}b^{\ell}_{\mu} - 
    \partial_{\mu}b^{\ell}_{\nu} + 
    g\varepsilon_{jk\ell}b^{j}_{\mu}b^{k}_{\nu}\; ,
    \label{eq:Fmunu}
\end{equation}
for the weak-isospin symmetry, and 
\begin{equation}
    f_{\mu\nu} = \partial_{\nu}{{\mathcal A}}_\mu - \partial_{\mu}{{\mathcal 
    A}}_\nu \; , 
    \label{eq:fmunu}
\end{equation}
for the weak-hypercharge symmetry.  We may summarize the interactions 
by the Lagrangian
\begin{equation}
\L = \L_{\rm gauge} + \L_{\rm leptons} \ ,                           
\end{equation}             
with
\begin{equation}
\L_{\rm gauge}=-\frac{1}{4}F_{\mu\nu}^\ell F^{\ell\mu\nu}
-\frac{1}{4}f_{\mu\nu}f^{\mu\nu},
\end{equation}
and
\begin{eqnarray}     
\L_{\rm leptons} & = & \overline{{\sf R}}\:i\gamma^\mu\!\left(\partial_\mu
+i\frac{g^\prime}{2}{\cal A}_\mu Y\right)\!{\sf R} 
\label{eq:matiere} \\ 
& + & \overline{{\sf
L}}\:i\gamma^\mu\!\left(\partial_\mu 
+i\frac{g^\prime}{2}{\cal
A}_\mu Y+i\frac{g}{2}\vec{\tau}\cdot\vec{b}_\mu\right)\!{\sf L}. \nonumber
\end{eqnarray}
The \ws\ gauge symmetry forbids a mass term for the electron in the 
matter piece \eqn{eq:matiere}.  Moreover, the theory we have described 
contains four massless electroweak gauge bosons, namely ${{\mathcal A}}_\mu$, 
$b^{1}_{\mu}$, $b^{2}_{\mu}$, and $b^{3}_{\mu}$, whereas Nature has 
but one: the photon.  To give masses to the gauge bosons and 
constituent fermions, we must hide the electroweak symmetry.

The most apt analogy for the hiding of the electroweak gauge 
symmetry is found in superconductivity.  In the Ginzburg-Landau 
description~\cite{4} of the superconducting phase transition, a 
superconducting material is regarded as a collection of two kinds of 
charge carriers: normal, resistive carriers, and superconducting, 
resistanceless carriers.

In the absence of a magnetic field, the free energy of the superconductor 
is related to the free energy in the normal state through
\begin{equation}
G_{\rm super}(0) = G_{\rm normal}(0) + \alpha \abs{\psi}^2 + \beta 
\abs{\psi}^4\;\;,
\end{equation}
where $\alpha$ and $\beta$ are phenomenological parameters and 
$\abs{\psi}^2$ is an order parameter that measures the density of 
superconducting charge carriers.  The parameter $\beta$ is non-negative, 
so that the free energy is bounded from below.

\begin{figure}
	\centerline{\BoxedEPSF{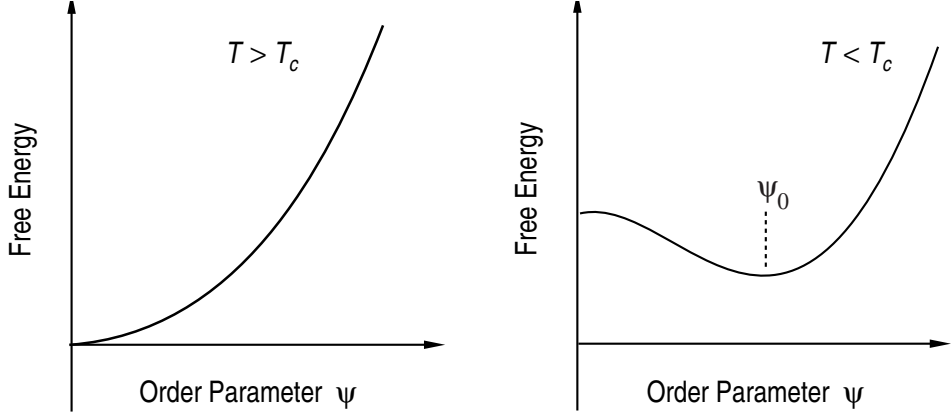  scaled 800}}
	\vspace*{6pt}
	\caption{Ginzburg-Landau description of the superconducting phase 
	transition.}
	\protect\label{fig1}
\end{figure}

Above the critical temperature for the onset of superconductivity, the 
parameter $\alpha$ is positive and the free energy of the substance is 
supposed to be an increasing function of the density of 
superconducting carriers, as shown in Figure \ref{fig1}(a).  The state 
of minimum energy, the vacuum state, then corresponds to a purely 
resistive flow, with no superconducting carriers active.  Below the 
critical temperature, the parameter $\alpha$ becomes negative and the 
free energy is minimized when $\psi = \psi_0 \ne 0$, as illustrated in 
Figure \ref{fig1}(b).

This is a nice cartoon description of the superconducting phase 
transition, but there is more.  In an applied magnetic field $\vec{H}$, 
the free energy is
\begin{equation}
G_{\rm super}(\vec{H}) = G_{\rm super}(0) + \frac{\vec{H}^2}{8\pi} + 
\frac{1}{2m^\star}|-i\hbar\nabla\psi-(e^\star/c)\vec{A}\psi|^2 
\;\;,
\end{equation}
where $e^\star$ and $m^\star$ are the charge ($-2$ units) and effective 
mass of the superconducting carriers.  In a weak, slowly varying field 
$\vec{H} \approx 0$, when we can approximate $\psi\approx\psi_0$ and 
$\nabla\psi\approx 0$, the usual variational analysis leads to the 
equation of motion,
\begin{equation}
\nabla^2\vec{A}-\frac{4\pi e^\star}{m^\star c^2}\abs{\psi_0}^2\vec{A} = 
0\;\;,
\end{equation}
the wave equation of a massive photon.  In other words, the photon 
acquires a mass within the superconductor.  This is the origin of the 
Meissner effect, the exclusion of a magnetic field from a 
superconductor.  More to the point for our purposes, it shows how a 
symmetry-hiding phase transition can lead to a massive gauge boson.

To give masses to the intermediate bosons of the weak interaction, we 
take advantage of a relativistic generalization of the Ginzburg-Landau 
phase transition known as the Higgs mechanism \cite{5}.  We introduce 
a complex doublet of scalar fields
\begin{equation}
\phi\equiv \left(\begin{array}{c} \phi^+ \\ \phi^0 \end{array}\right)
\end{equation}
with weak hypercharge $Y_\phi=+1$.  Next, we add to the Lagrangian new 
(gauge-invariant) terms for the interaction and propagation of the 
scalars,
\begin{equation}
      \L_{\rm scalar} = (\D^\mu\phi)^\dagger(\D_\mu\phi) - V(\phi^\dagger \phi),
\end{equation}
where the gauge-covariant derivative is
\begin{equation}
      \D_\mu=\partial_\mu 
+i\frac{g^\prime}{2}{\cal A}_\mu
Y+i\frac{g}{2}\vec{\tau}\cdot\vec{b}_\mu \; ,
\label{eq:GcD}
\end{equation}
and the potential interaction has the form
\begin{equation}
      V(\phi^\dagger \phi) = \mu^2(\phi^\dagger \phi) +
\abs{\lambda}(\phi^\dagger \phi)^2 .
\label{SSBpot}
\end{equation}
We are also free to add a Yukawa interaction between the scalar fields
and the leptons,
\begin{equation}
      \L_{\rm Yukawa} = -G_e\left[\overline{{\sf R}}(\phi^\dagger{\sf
L}) + (\overline{{\sf L}}\phi){\sf R}\right].
\label{eq:Yukterm}
\end{equation}

We then arrange 
their self-interactions so that the vacuum state corresponds to a 
broken-symmetry solution.  The electroweak symmetry is spontaneously broken if the parameter
$\mu^2<0$. The minimum energy, or vacuum state, may then be chosen
to correspond to the vacuum expectation value
\begin{equation}
\vev{\phi} = \left(\begin{array}{c} 0 \\ v/\sqrt{2} \end{array}
\right),
\label{eq:vevis}
\end{equation}
where the numerical value of
\begin{equation}
      v = \sqrt{-\mu^2/\abs{\lambda}} =
\left(G_F\sqrt{2}\right)^{-\frac{1}{2}}  \approx 246\gev 
\label{eq:vevdef}
\end{equation}
is fixed by the low-energy phenomenology of charged-current
interactions.

As a result of spontaneous symmetry breaking, the weak bosons acquire 
masses, as auxiliary scalars assume the role of the third 
(longitudinal) degrees of freedom of what had been massless gauge 
bosons.  Specifically, the mediator of the charged-current weak 
interaction, $W^{\pm} = (b_{1} \mp ib_{2})/\sqrt{2}$, acquires a 
mass characterized by 
$M_W^2=\pi\alpha/G_F\sqrt{2}\sin^2{\theta_W}$, where $\theta_W$ is the 
weak mixing angle. The mediator of the neutral-current weak 
interaction, $Z = b_{3}\cos{\theta_{W}} - \mathcal{A}\sin{\theta_{W}}$, 
acquires a mass characterized by 
$M_Z^2=M_W^2/\cos^2{\theta_W}$.  After spontaneous symmetry breaking, 
there remains an unbroken $U(1)_{\mathrm{em}}$ phase symmetry, so that 
electromagnetism is mediated by a massless photon, $A = 
\mathcal{A}\cos{\theta_{W}} + b_{3}\sin{\theta_{W}}$, coupled to the 
electric charge $e = gg^{\prime}/\sqrt{g^{2} + g^{\prime 2}}$.  As a vestige 
of the spontaneous breaking of the symmetry, there remains a massive, 
spin-zero particle, the Higgs boson.  The mass of the Higgs scalar is 
given symbolically as $M_{H}^{2} = -2\mu^{2} > 0$, but we have no 
prediction for its value.  Though what we take to be the work of the 
Higgs boson is all around us, the Higgs particle itself has not yet 
been observed.

The fermions (the electron in our abbreviated treatment) acquire 
masses as well; these are determined not only by the scale of 
electroweak symmetry breaking, $v$, but also by  their Yukawa interactions with
the scalars.  The mass of the electron is set by the dimensionless 
coupling constant $G_{e} = m_{e}\sqrt{2}/v \approx 3 \times 10^{-6}$, 
which is both small and---so far as we now know---arbitrary.
\subsection{Experimental Update}
It will be helpful for orientation to recall some of the recent 
precision electroweak measurements as presented at the DPF99 
Conference in Los Angeles \cite{DPF99,Marciano:1999ia}.  We will go 
looking for trouble in \S\ref{sub:issues} below, but the overall 
assessment is that electroweak observables are in accord with the 
\begin{figure}
	\centerline{\BoxedEPSF{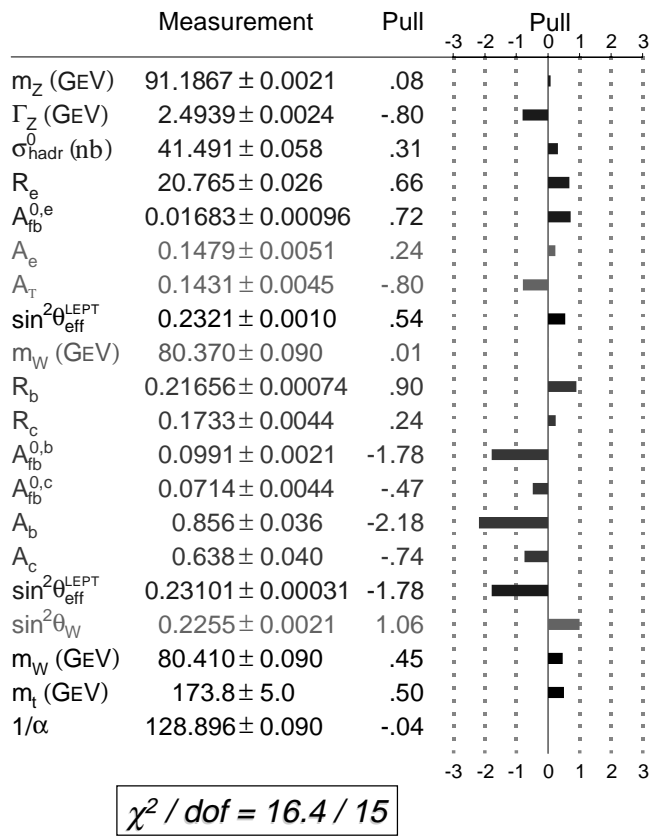  scaled 500}}
	\vspace*{6pt}
	\caption{Precision electroweak measurements and the pulls they exert 
	on a global fit to the standard model, from {\protect{Ref.\ \cite{karlen}}}.}
	\protect\label{fig:pulls}
\end{figure}
predictions of the standard model at the level of 0.1\%\ 
\cite{karlen,erlerlang}.  The degree of agreement is summarized 
pictorially in Figure \ref{fig:pulls} (\cf\ Table 1 of Ref.\ 
\cite{erlerlang}).  Taken together, the $Z^{0}$-pole 
data from the LEP experiments and SLD yield a weak mixing parameter
\begin{equation}
	\sin^{2}\theta_{W}^{\mathrm{eff}} = 0.23128 \pm 0.00022.
	\label{eq:xWval}
\end{equation}
Direct measurements at LEP~2 and the Tevatron give the $W$-boson mass
\begin{equation}
	M_{W} = (80.39 \pm 0.06)\gevcc \; ,
	\label{eq:MWval}
\end{equation}
while the ``world-average'' top-quark mass from CDF and D\O\ is
\begin{equation}
	m_{t} = (174.3 \pm 5.1)\gevcc \; .
	\label{eq:mtval}
\end{equation}
The NuTeV experiment at Fermilab has reported a competitive indirect 
determination of the $W$ mass, inferred from measurements of the 
$\nu_{\mu}N$ and $\bar{\nu}_{\mu}N$ cross sections.  They find
\begin{equation}
	M_{W} = (80.26 \pm 0.11)\gevcc \; .
	\label{eq:nuMWval}
\end{equation}
Thanks to a new evaluation of the finite part of the 
$\mathcal{O}(\alpha^{2})$ correction to the muon lifetime 
\cite{stuvanR}, we have a new determination of the Fermi constant 
measured in muon decay, 
\begin{equation}
	G_{\mu} = (1.16637 \pm 0.00001)\times 10^{-5}\gev^{-2}\; .
	\label{eq:Gmu}
\end{equation}

Bennett and Wieman (Boulder) have reported a new determination of the 
weak charge of Cesium by measuring the transition polarizability for the 
6S-7S transition \cite{boulder}.  The new value,
\begin{equation}
	Q_W(\textrm{Cs})= -72.06 \pm 0.28\hbox{ (expt)} \pm 0.34\hbox{ 
	(theory),} 
	\label{eq:wieman}
\end{equation}
represents a seven-fold improvement in the experimental error and a 
significant reduction in the theoretical uncertainty.  It differs by 
2.5 standard deviations from the prediction of the standard model.  We 
are left with the traditional situation in which elegant measurements 
of parity nonconservation in atoms are on the edge of incompatibility 
with the standard model.

From the wealth of particle searches and cross-section measurements at 
LEP~2, let us simply remark that no anomalies whatever have been noted 
in the reactions
\begin{equation}
	e^{+}e^{-} \to \left\{ 
	\begin{array}{c}
		W^{+}W^{-}  \\
		Z^{0}Z^{0}  \\
		\ell^{+}\ell^{-}  \\
		q\bar{q}
	\end{array}\right. \; .
	\label{eq:searches}
\end{equation}
Similarly, the overall conclusion from HERA is that the 
neutral-current and charged-current cross sections measured in 
$e^{+}p$ collisions have the expected character and reproduce the 
known values of $M_{W}$ and $M_{Z}$.

\subsection{Experimental Clues about $M_{H}$}
The success of the electroweak theory means that it makes sense to 
use standard-model fits to the electroweak observables to determine 
``best'' values for the parameters that are not yet directly 
constrained by experiment.  Over the past decade, the greatest 
sensitivity has been to the value of the top-quark mass, and fits to 
the electroweak observables gave early indications for the great mass 
of the top quark \cite{CQPT}.  Now that the top-quark mass is known 
rather well from Tevatron experiments, we can interrogate the quantum 
corrections to electroweak observables for the best value of the 
Higgs-boson mass.  In detail, the inferences depend upon the data set 
selected and the values adopted for the ``known'' parameters, 
including the value of the fine structure constant 
$\alpha(M_{Z}^{2})$ evaluated at the $Z^{0}$ pole.  The consensus of 
the fits is that, \textit{within the standard electroweak theory,} 
the Higgs boson may be just around the corner.  In the global fit of 
Erler and Langacker \cite{erlerlang}, which is representative of 
other work, the best-fit value for the mass of the standard-model 
Higgs boson is 
\begin{equation}
	M_H = 107^{+67}_{-45}\gevcc \; ,	
	\label{eq:Hfit}
\end{equation}
and the 95\% CL upper limit is $M_{H} \ltap 255\gevcc$. A very 
interesting question is, how are these constraints relaxed in specific 
theories other than the standard model?

\subsection{Some Experimental Issues  {\protect{\label{sub:issues}}}}
Suppose, in the face of the spectacular successes of the electroweak 
theory, we go looking for trouble.  Where might we find it?  The 
heavy top quark gives rise to the theoretical suspicion that 
anomalies are most likely to show themselves in the third generation 
of quarks and leptons.  As it happens, the only suggestive anomaly in precision 
measurements on the $Z^{0}$ pole involves $b$ quarks.  The 
forward-backward asymmetry for $b\bar{b}$ events measured at LEP and 
the left-right forward-backward asymmetry for $b\bar{b}$ events measured 
at SLD indicate a three-standard-deviation difference from the 
standard model for 
\begin{equation}
	A_{b} = \frac{L_{b}^{2} - R_{b}^{2}}{L_{b}^{2} + R_{b}^{2}} \; ,
	\label{eq:Abdef}
\end{equation}
where $L_{b}$ and $R_{b}$ are the left-handed and right-handed chiral 
couplings of the $Z$ to $b$ quarks.  At tree level in the standard 
model, they take the values
\begin{eqnarray}
	L_{b}^{\mathrm{theory}} = & -1 + \cfrac{2}{3}\sin^{2}\theta_{W} & 
	\approx -0.846 \;\; ,
	\label{eq:chivals}  \\
	R_{b}^{\mathrm{theory}} = & \cfrac{2}{3}\sin^{2}\theta_{W} & \approx 
	0.154 \;\; ,\nonumber
\end{eqnarray}
Current measurements imply that
\begin{equation}
	A_{b}^{\mathrm{exp}} = (0.94 \pm 0.02)A_{b}^{\mathrm{theory}} \;\; .
	\label{eq:Abrat}
\end{equation}

We must reconcile this apparent discrepancy with the good agreement 
between the quantity $R_{b} = \Gamma(Z^{0} \to 
b\bar{b})/\Gamma(Z^{0} \to\hbox{hadrons})$, which is sensitive to the 
combination $L_{b}^{2} + R_{b}^{2}$.  The current data say that
\begin{equation}
	R_{b}^{\mathrm{exp}} = (1.004 \pm 0.004)R_{b}^{\mathrm{theory}} \; ,
	\label{eq:Rbrat}
\end{equation}
which implies that
\begin{equation}
	L_{b}^{2} + R_{b}^{2} = 0.7432 \pm 0.0040 \; .
	\label{eq:chiralsum}
\end{equation}
We can solve \eqn{eq:Abrat} and \eqn{eq:chiralsum} simultaneously; choosing the 
appropriate signs, we find 
\begin{eqnarray}
	L_{b}^{\mathrm{exp}} & = & -0.836 \pm 0.004 \; ,
	\label{eq:Expchi}  \\
	R_{b}^{\mathrm{exp}} & = & 0.2117 \pm 0.0176 \; .
	\nonumber
\end{eqnarray}
Expressed as deviations from the standard model, we have 
\begin{eqnarray}
	\delta L_{b} & \equiv L_{b}^{\mathrm{theory}} - 
	L_{b}^{\mathrm{exp}} & = -0.010 \pm 0.004 \; ,
	\label{eq:chidiffs}  \\
	\delta R_{b} & \equiv R_{b}^{\mathrm{theory}} - 
	R_{b}^{\mathrm{exp}} & = -0.0577 \pm 0.0176 \; .
	\nonumber
\end{eqnarray}
Neither effect is titanic!  However, the suggestion that $\delta 
R_{b}/R_{b}^{\mathrm{theory}} \approx -40\%$ (whereas $\delta 
L_{b}/L_{b}^{\mathrm{theory}} \approx 1\%$) can be taken as an 
indication that if we want to look for trouble, the right-handed $b$ 
coupling is the place to look.  If this anomaly is real, we might 
expect to observe flavor-changing neutral-current transitions $b \to 
s$, $b \to d$, and $s \to d$.

\subsection{An Assessment}
Experiments over the past twenty-five years have brought us numerous 
confirmations of the \ws\ electroweak theory: the existence of 
neutral currents, the necessity of charm, and the existence and 
properties of the weak gauge bosons $W^{\pm}$ and $Z^{0}$.  Experiment 
has also given essential guidance to the form of the evolving standard 
model through the discovery of the third generation of leptons 
$(\nu_{\tau},\tau)$ and quarks $(t,b)$.  And, finally, experiment has 
given us a number of significant surprises that have shaped both 
experimental and theoretical opportunities: the narrowness of $J\!/\!\psi$ 
and $\psi^{\prime}$, the unexpectedly long $B$ lifetime, the large 
degree of $B^{0}$--$\bar{B}^{0}$ mixing, the extreme heaviness of the 
top quark, and---very likely---evidence of neutrino oscillations.

Ten years of precision measurements have found no significant 
deviations from the predictions of the electroweak theory.  A series 
of quite remarkable experiments, not to mention the accompanying 
evolution in theorectical calculations, have tested the quantum corrections of 
the electroweak theory---loop effects---to a precision of one per 
mil.  The net result of this prodigious effort is that we have no 
found no evidence for new physics \ldots yet.

It is remarkable that the resulting theory has been tested at distances 
ranging from about $10^{-17}~\hbox{cm}$ to about $4\times 
10^{20}~\hbox{cm}$, especially when we consider that classical 
electrodynamics has its roots in the tabletop experiments that gave us 
Coulomb's law.  These basic ideas were modified in response to the 
quantum effects observed in atomic experiments.  High-energy physics 
experiments both inspired and tested the unification of weak and 
electromagnetic interactions.  At distances longer than common 
experience, electrodynamics---in the form of the statement that the 
photon is massless---has been tested in measurements of the magnetic 
fields of the planets.  With additional assumptions, the observed 
stability of the Magellanic clouds provides evidence that the photon is 
massless over distances of about $10^{22}~\hbox{cm}$ \cite{gammamass}.

The extraordinary success of the electroweak theory leaves us with 
these urgent questions: Is the electroweak theory true?  Can it be 
complete?

\section{The Standard-Model Higgs Boson}
\subsection{Why the Higgs Boson Must Exist}
\begin{figure}[tb]
	\centerline{\BoxedEPSF{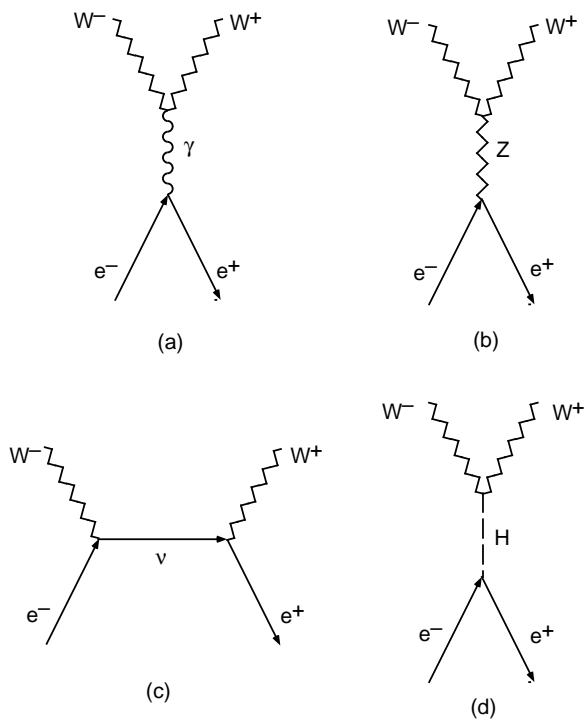  scaled 700}}
	\vspace*{6pt}
	\caption{Lowest-order contributions to the $e^+e^- \rightarrow 
	W^{+}W^{-}$ scattering amplitude.}
	\protect\label{fig:eeWW}
\end{figure}

How can we be sure that a Higgs boson, or something very like it, will be 
found? One 
path to the \emph{theoretical} discovery of the Higgs boson
involves its role in the cancellation of 
high-energy divergences. An illuminating example is provided by the 
reaction
\begin{equation}
	e^+e^- \to W^+W^-,
\end{equation}
which is described in lowest order by the four 
Feynman graphs in Figure \ref{fig:eeWW}. The contributions of the direct-channel 
$\gamma$- and $Z^0$-exchange diagrams 
of Figs.~\ref{fig:eeWW}(a) and (b) cancel the leading divergence in the $J=1$ 
partial-wave amplitude of 
the neutrino-exchange diagram in Figure~\ref{fig:eeWW}(c).  This is 
the famous ``gauge cancellation'' observed in experiments at LEP~2 
and the Tevatron.

However, the $J=0$ partial-wave amplitude, which exists in this 
case because 
the electrons are massive and may therefore be found in the ``wrong'' 
helicity state, grows as $s^{1/2}$ for the production of longitudinally 
polarized gauge bosons. The resulting divergence is precisely cancelled by 
the Higgs boson graph of Figure~\ref{fig:eeWW}(d). If the Higgs boson did not exist, 
something else would have to play this role. From the point of view 
of $S$-matrix analysis, the Higgs-electron-electron coupling must be 
proportional to the electron mass, because ``wrong-helicity'' amplitudes 
are always proportional to the fermion mass.

Let us underline this result.
If the gauge symmetry were unbroken, there would be 
no Higgs boson, no longitudinal gauge bosons, and no extreme divergence 
difficulties. But there would be no viable low-energy phenomenology
 of the 
weak interactions. The most severe divergences of individual diagrams 
are eliminated by the gauge 
structure of the couplings among gauge bosons and leptons. A lesser, but 
still potentially fatal, divergence arises because the electron has 
acquired mass---because of the Higgs mechanism. Spontaneous symmetry 
breaking provides its own cure by supplying a Higgs boson to remove the 
last divergence. A similar interplay and compensation must exist in any 
satisfactory theory.

\subsection{Bounds on $M_{H}$}
The Standard Model does not give a precise 
prediction for the mass of the Higgs boson. We can, however, use arguments 
of self-consistency to place plausible lower and upper bounds on the mass of 
the Higgs particle in the minimal model. Unitarity arguments~\cite{lqt} lead to a conditional upper bound on the Higgs 
boson mass. It is straightforward to compute the 
amplitudes ${\cal M}$ for gauge boson scattering at high energies, and to make
a partial-wave decomposition, according to
\beq
      {\cal M}(s,t)=16\pi\sum_J(2J+1)a_J(s)P_J(\cos{\theta}) \; .
\eeq
 Most channels ``decouple,'' in the sense 
that partial-wave amplitudes are small at all energies (except very
near the particle poles, or at exponentially large energies), for
any value of the Higgs boson mass $M_H$. Four channels are interesting:
\beq
\begin{array}{cccc}
W_L^+W_L^- & Z_L^0Z_L^0/\sqrt{2} & HH/\sqrt{2} & HZ_L^0 \; ,
\end{array}
\eeq
where the subscript $L$ denotes the longitudinal polarization
states, and the factors of $\sqrt{2}$ account for identical particle
statistics. For these, the $s$-wave amplitudes are all asymptotically
constant (\ie, well-behaved) and  
proportional to $G_FM_H^2.$ In the high-energy 
limit,\footnote{It is convenient to calculate these amplitudes by 
means of the Goldstone-boson equivalence theorem \cite{EQT}, which 
reduces the dynamics of longitudinally polarized gauge bosons to a 
scalar field theory with interaction Lagrangian given by 
$\mathcal{L}_{\mathrm{int}} = -\lambda v h 
(2w^{+}w^{-}+z^{2}+h^{2}) - 
(\lambda/4)(2w^{+}w^{-}+z^{2}+h^{2})^{2}$, with $1/v^{2} = 
G_{F}\sqrt{2}$ and $\lambda = G_{F}M_{H}^{2}/\sqrt{2}$.}
\beq
\lim_{s\gg M_H^2}(a_0)\to\frac{-G_F M_H^2}{4\pi\sqrt{2}}\cdot \left[
\begin{array}{cccc} 1 & 1/\sqrt{8} & 1/\sqrt{8} & 0 \\
      1/\sqrt{8} & 3/4 & 1/4 & 0 \\
      1/\sqrt{8} & 1/4 & 3/4 & 0 \\
      0 & 0 & 0 & 1/2 \end{array} \right] \; .
\eeq 
Requiring that the largest eigenvalue respect the 
partial-wave unitarity condition $\abs{a_0}\le 1$ yields
\beq
	M_H \le \left(\frac{8\pi\sqrt{2}}{3G_F}\right)^{1/2} =1\tevcc
\eeq
as a condition for perturbative unitarity.

If the bound is respected, weak interactions remain weak at all
energies, and perturbation theory is everywhere reliable. If the
bound is violated, perturbation theory breaks down, and weak
interactions among $W^\pm$, $Z$, and $H$ become strong on the \onetev.
This means that the features of strong interactions at GeV energies
will come to characterize electroweak gauge boson interactions at
TeV energies. We interpret this to mean that new phenomena are to
be found in the electroweak interactions at energies not much larger
than 1~TeV.

It is worthwhile to note in passing that 
the threshold behavior of the partial-wave amplitudes for gauge-boson 
scattering follows generally from chiral symmetry \cite{LT8}.  The partial-wave 
amplitudes $a_{IJ}$ of definite isospin $I$ and angular momentum $J$ are 
given by
\begin{eqnarray}
	a_{00} \approx & G_Fs/8\pi\sqrt{2} & \hbox{attractive,}
	\nonumber \\
	a_{11}  \approx & G_Fs/48\pi\sqrt{2} & \hbox{attractive,} \\
	a_{20} \approx & -G_Fs/16\pi\sqrt{2} & \hbox{repulsive.}
	\nonumber
\end{eqnarray} 

The electroweak theory itself provides another reason to expect that 
discoveries will not end with the Higgs boson.  Scalar field theories 
make sense on all energy scales only if they are noninteracting, or 
``trivial'' \cite{15}.  The vacuum of quantum field theory is a dielectric 
medium that screens charge.  Accordingly, the effective charge is a 
function of the distance or, equivalently, of the energy scale.  This is 
the famous phenomenon of the running coupling constant.

In $\lambda\phi^4$ theory (compare the interaction term in the Higgs 
potential), it is easy to calculate the variation of the coupling 
constant $\lambda$ in perturbation theory by summing bubble graphs like 
this one:
\begin{equation}
\BoxedEPSF{Bulle.epsf  scaled 600}\;\;\;\;.
\end{equation} \vphantom{{\LARGE |}}The coupling constant $\lambda(\mu)$ on a physical scale $\mu$ 
is related 
to the coupling constant on a higher scale $\Lambda$ by
\begin{equation}
\frac{1}{\lambda(\mu)} = \frac{1}{\lambda(\Lambda)} + 
\frac{3}{2\pi^2}\log{\left(\Lambda/\mu\right)}\;\;.
\label{rng}
\end{equation}
This perturbation-theory result is reliable only when $\lambda$ is small, 
but lattice field theory allows us to treat the strong-coupling regime.

In order for the Higgs potential to be stable (\ie, for the energy of the 
vacuum state not to race off to $-\infty$), $\lambda(\Lambda)$ must not 
be negative.  Therefore we can rewrite \eqn{rng} as an inequality,
\begin{equation}
\frac{1}{\lambda(\mu)} \ge 
\frac{3}{2\pi^2}\log{\left(\Lambda/\mu\right)}\;\;. 
\end{equation}
This gives us an {\em upper bound},
\begin{equation}
\lambda(\mu) \le 
2\pi^2/3\log{\left(\Lambda/\mu\right)}\;\;,
\label{upb}
\end{equation}
on the coupling strength at the physical scale $\mu$.
If we require the theory to make sense to arbitrarily high energies---or 
short distances---then we must take the limit $\Lambda\rightarrow\infty$ 
while holding $\mu$ fixed at some reasonable physical scale.  In this 
limit, the bound \eqn{upb} forces $\lambda(\mu)$ to zero.  The scalar field 
theory has become free field theory; in theorist's jargon, it is trivial.

We can rewrite the inequality \eqn{upb} as a bound on the Higgs-boson mass.  
Rearranging and exponentiating both sides gives the condition
\begin{equation}
\Lambda \le \mu \exp{\left(\frac{2\pi^2}{3\lambda(\mu)}\right)}\;\;.
\end{equation}
Choosing the physical scale as $\mu=M_H$, and remembering that, before 
quantum corrections,
\begin{equation}
M_H^2 = 2\lambda(M_H)v^2\;\;,
\end{equation}
where $v=(G_F\sqrt{2})^{-1/2}\approx 246~\hbox{GeV}$ is the vacuum 
expectation value of the Higgs field times $\sqrt{2}$, we find that
\begin{equation}
\Lambda \le M_H\exp{\left(\frac{4\pi^2v^2}{3M_H^2}\right)}\;\;.
\end{equation}
For any given Higgs-boson mass, there is a maximum energy scale 
$\Lambda^\star$ at which the theory ceases to make sense.  The 
description of the Higgs boson as an elementary scalar is at best an 
effective theory, valid over a finite range of energies.

\begin{figure}[tb]
	\centerline{\BoxedEPSF{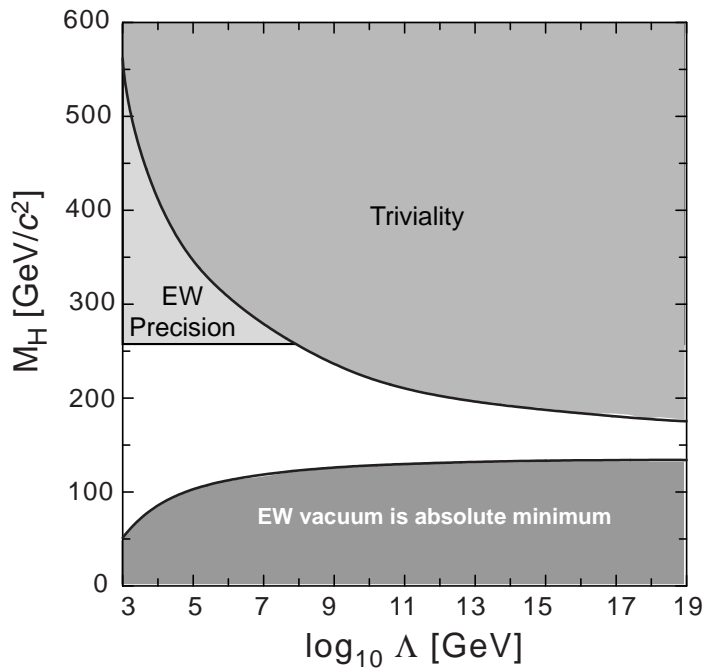  scaled 700}}
	\vspace*{6pt}
	\caption{Bounds on the Higgs-boson mass that follow from 
	requirements that the electroweak theory be consistent up to the 
	energy scale $\Lambda$.  The upper bound follows from triviality 
	conditions; the lower bound follows from the requirement that $V(v) < 
	V(0)$.  Also shown is the range of masses permitted at the 95\%\ 
	confidence level by precision measurements.}
	\protect\label{fig:Hbds}
\end{figure}

This perturbative analysis breaks down when the Higgs-boson mass 
approaches $1~\hbox{TeV}\!/\!c^2$ and the interactions become strong.  
Lattice analyses~\cite{17} indicate that, for the theory to describe 
physics to an accuracy of a few percent up 
to a few TeV, the mass of the Higgs boson can be no more than about 
$710\pm 60\gevcc$.  Another way of putting this result is that, if 
the elementary Higgs boson takes on the largest mass allowed by 
perturbative unitarity arguments, the electroweak theory will be living 
on the brink of instability.

A lower bound is obtained by 
computing~\cite{SSI18} the first quantum corrections to the classical potential
\eqn{SSBpot}. Requiring that $\vev{\phi}\neq 0$ be an absolute minimum of the one-loop 
potential up to a scale $\Lambda$ yields the vacuum-stability condition 
\begin{equation}
	M_H^2 > \frac{3G_F\sqrt{2}}{8\pi^{2}}(2M_W^4+M_Z^4-4m_{t}^{4})
	\log(\Lambda^{2}/v^{2}) \; .
\end{equation}

The upper and lower bounds plotted in Figure \ref{fig:Hbds} are the results of 
full two-loop calculations \cite{2loopvacstab}.  There I have also 
indicated the upper bound on $M_{H}$ derived from precision 
electroweak measurements \textit{in the framework of the standard electroweak 
theory.}  If the Higgs boson is relatively light---which would itself require 
explana\-tion---then the theory can be self-consistent up to 
very high energies.  If the electroweak theory is to make sense all the 
way up to a unification scale $\Lambda^\star = 10^{16}~\hbox{GeV}$, then 
the Higgs-boson mass must lie in the interval $145\gevcc \ltap M_{W} \ltap 170
\gevcc$ \cite{16}.

\subsection{Higgs-Boson Properties}
Once we assume a value for the Higgs-boson mass, it is a simple matter 
to compute the rates for Higgs-boson decay into pairs of fermions or 
weak bosons \cite{Ellis:1976ap}.  For a fermion with color $N_{c}$, the partial width is
\begin{equation}
	\Gamma(H \to f\bar{f}) = \frac{G_{F}m_{f}^{2}M_{H}}{4\pi\sqrt{2}} 
	\cdot N_{c} \cdot \left( 1 - \frac{4m_{f}^{2}}{M_{H}^{2}} 
	\right)^{3/2} \; ,
	\label{eq:Higgsff}
\end{equation}
which is proportional to $M_{H}$ in the limit of large Higgs mass.
The partial width for decay into a $W^{+}W^{-}$ pair is
\begin{equation}
	\Gamma(H \to W^{+}W^{-}) = \frac{G_{F}M_{H}^{3}}{32\pi\sqrt{2}} 
	(1 - x)^{1/2} (4 -4x +3x^{2}) \; ,
	\label{eq:HiggsWW}
\end{equation}
where $x \equiv 4M_{W}^{2}/M_{H}^{2}$.  Similarly, the partial width 
for decay into a pair of $Z^{0}$ bosons is 
\begin{equation}
	\Gamma(H \to Z^{0}Z^{0}) = \frac{G_{F}M_{H}^{3}}{64\pi\sqrt{2}} 
	(1 - x^{\prime})^{1/2} (4 -4x^{\prime} +3x^{\prime 2}) \; ,
	\label{eq:HiggsZZ}
\end{equation}
where $x^{\prime} \equiv 4M_{Z}^{2}/M_{H}^{2}$.  The rates for decays into 
weak-boson pairs are asymptotically proportional to $M_{H}^{3}$ and 
$\cfrac{1}{2}M_{H}^{3}$, respectively, the factor $\cfrac{1}{2}$ 
arising from weak isospin.  In the final factors of \eqn{eq:HiggsWW} 
and \eqn{eq:HiggsZZ}, $2x^{2}$ and $2x^{\prime 2}$, respectively, 
arise from decays into transversely polarized gauge bosons.  The 
dominant decays for large $M_{H}$ are into pairs of longitudinally 
polarized weak bosons.

\begin{figure}[t!]
	\centerline{\BoxedEPSF{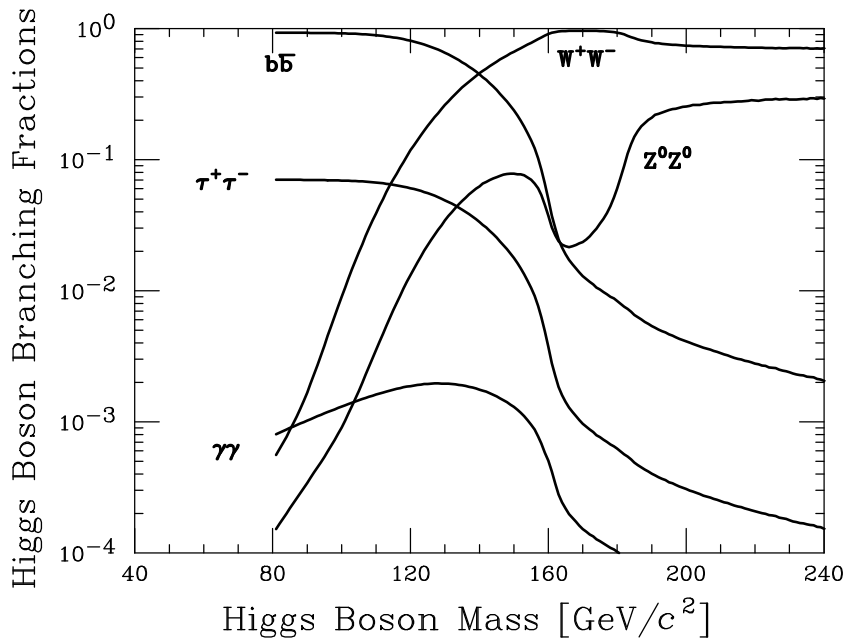  scaled 1000}}
	\vspace*{6pt}
	\caption{Branching fractions for the prominent decay modes of a light 
	Higgs boson.}
	\protect\label{fig:LHdk}
\end{figure}

Branching fractions for decay modes that hold promise for the 
detection of a light Higgs boson are displayed in Figure 
\ref{fig:LHdk}.  In addition to the $f\bar{f}$ and $VV$ modes that 
arise at tree level, I have included the $\gamma\gamma$ mode that 
proceeds through loop diagrams.  Though rare, the $\gamma\gamma$ 
channel offers an important target for LHC experiments.

\begin{figure}[t!]
	\centerline{\BoxedEPSF{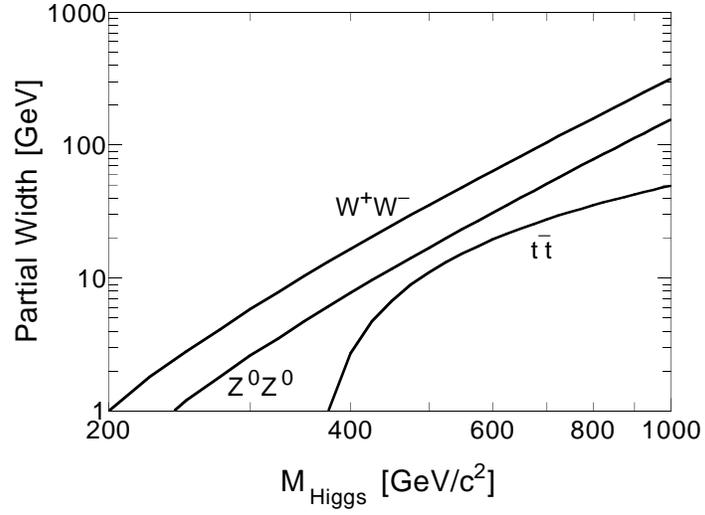  scaled 800}}
	\vspace*{6pt}
	\caption{Partial widths for the prominent decay modes of a heavy Higgs 
	boson.}
	\protect\label{fig:HHdk}
\end{figure}

Figure \ref{fig:HHdk} shows the partial widths for the decay of a 
Higgs boson into the dominant $W^+W^-$ and $Z^0Z^0$ channels and into 
$t\bar{t}$, for $m_t = 175\gevcc$.  Whether the $t\bar{t}$ mode will 
be useful to confirm the observation of a heavy Higgs boson, or merely 
drains probability from the $ZZ$ channel favored for a heavy-Higgs 
search, is a question for detailed detector simulations.

\begin{figure}[t!]
	\centerline{\BoxedEPSF{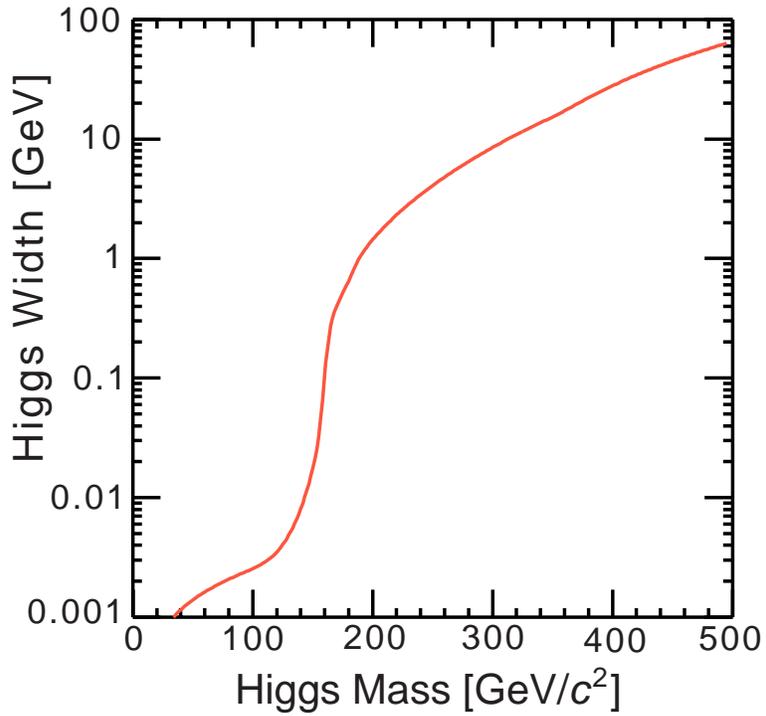  scaled 1300}}
	\vspace*{6pt}
	\caption{Higgs-boson total width as a function of mass.}
	\protect\label{fig:Htot}
\end{figure}

Below the $W^{+}W^{-}$ threshold, the total width of the 
standard-model Higgs boson is rather small, typically less than 
$1\gev$.  Far above the threshold for decay into gauge-boson pairs, 
the total width is proportional to $M_{H}^{3}$.  At masses 
approaching $1\tevcc$, the Higgs boson is an ephemeron, with a 
perturbative width approaching its mass.  The Higgs-boson total width 
is plotted as a function of $M_{H}$ in Figure \ref{fig:Htot}.

\subsection{Higgs-Boson Searches {\protect{\cite{hhg}}}}
\subsubsection{$e^{+}e^{-}$ Collisions at LEP}
\begin{figure}[tb]
	\centerline{\BoxedEPSF{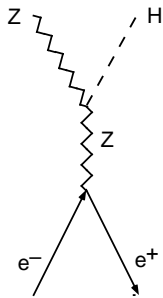  scaled 700}}
	\vspace*{6pt}
	\caption{Lowest-order contributions to the $e^+e^- \rightarrow 
	HZ^{0}$ scattering amplitude.}
	\protect\label{fig:eeHZ}
\end{figure}
Because the standard-model Higgs boson couples to fermion mass, the 
cross section for the reaction $e^{+}e^{-} \to H \to \hbox{all}$ is 
minute ($\propto m_{e}^{2}$).  With any remotely conceivable 
luminosity, even the narrowness of a light Higgs boson is not enough 
to make it visible.  This circumstance sets aside a traditional 
strength of electron-positron colliders: pole physics.  Instead, the 
most promising reaction for Higgs-boson physics at an $e^{+}e^{-}$ 
collider is associated production,
\begin{equation}
	e^{+}e^{-} \to HZ \; ,
	\label{eq:eeHZ}
\end{equation}
that corresponds to the Feynman diagram in Figure \ref{fig:eeHZ}, 
which has no small couplings.  The cross section \cite{sigHZ},
\begin{equation}
	\sigma = \frac{\pi\alpha^{2}}{8\sqrt{s}} 
	\frac{K(K^{2}+3M_{Z}^{2})[1 + (1-4x_{W})^{2}]}{(s-M_{Z}^{2})^{2} 
	\;\;\;x_{W}^{2}(1-x_{W})^{2}} \; ,
	\label{eq:sigHZ}
\end{equation}
where $K$ is the c.m.\ momentum of the Higgs boson and $x_{W}\equiv 
\sin^{2}\theta_{W}$, approaches about ten percent of 
$\sigma(e^{+}e^{-} \to \mu^{+}\mu^{-})$.

Searching in the observable channels of the reaction \eqn{eq:eeHZ}, 
the four LEP~2 experiments are sensitive nearly to the kinematical 
boundary
\begin{equation}
	M_{H}^{\mathrm{max}} = \sqrt{s} - M_{Z}\; .
	\label{eq:kinlim}
\end{equation}
Recent running at $\sqrt{s} = 189\gev$ leads to upper limits that lie 
within a few$\gevcc$ of $M_{H}^{\mathrm{max}}$ \cite{Hsearch}.  If 
the Higgs boson is established at LEP~2, it should be possible to 
determine its mass within a few hundred$\mevcc$ \cite{gunion}.
%%%%%%%%%%%%%%%%%%%%%%%%%%%%%%%%%%%%%%%%%%%%%%%%%%%%%%%%%%%%%%%%%%%%%%%%%%%%%%
%                                                                            %
%   The associated-production cross section given by \eqn{eq:sigHZ} is       %
%   shown in Figure \ref{fig:eeHZvsE} for Higgs-boson masses of 90, 100,     %
%   and $110\gevcc$.  These curves do not include the important effect of    %
%   initial-state radiation \cite{isrref}.                                   %
%                                                                            %
%%%%%%%%%%%%%%%%%%%%%%%%%%%%%%%%%%%%%%%%%%%%%%%%%%%%%%%%%%%%%%%%%%%%%%%%%%%%%%

Well above threshold, the angular distribution of Higgs production is 
characteristic of the $CP$ character of the Higgs boson.  For the 
$CP$-even standard-model Higgs boson,
\begin{equation}
	\frac{1}{\sigma}\frac{d\sigma}{d\cos\theta} \propto \sin^{2}\theta \; ,
	\label{eq:cpeven}
\end{equation}
while for a $CP$-odd Higgs boson, the production angular distribution 
is 
\begin{equation}
	\frac{1}{\sigma}\frac{d\sigma}{d\cos\theta} \propto 1 + \cos^{2}\theta \; .
	\label{eq:cpodd}
\end{equation}
The angular distribution will be a powerful diagnostic---once a Higgs 
boson is observed.  For other techniques to determine the parity of a 
Higgs particle, see \cite{Kramer:1994jn}.

%%%%%%%%%%%%%%%%%%%%%%%%%%%%%%%%%%%%%%%%%%%%%%%%%%%%%%%%%%%%%%%%%%%%%%%%%%%%%
%                                                                           %
%   \begin{figure}[tb]                                                      %
%       \centerline{\BoxedEPSF{eeHZvsE.eps  scaled 700}}                    %
%       \vspace*{6pt}                                                       %
%       \caption{Energy dependence of the cross section for associated      %
%       production, $e^+e^- \rightarrow HZ^{0}$, for three values of the    %
%       Higgs-boson mass.}                                                  %
%       \protect\label{fig:eeHZvsE}                                         %
%   \end{figure}                                                            %
%                                                                           %
%%%%%%%%%%%%%%%%%%%%%%%%%%%%%%%%%%%%%%%%%%%%%%%%%%%%%%%%%%%%%%%%%%%%%%%%%%%%%
\subsubsection{$e^{+}e^{-}$ Collisions at a Linear Collider}
At $e^{+}e^{-}$ linear colliders above the LEP~2 energy scale, the 
most promising reactions for the production of Higgs bosons are 
\eqn{eq:eeHZ} plus the gauge-boson--fusion reactions
\begin{equation}
	e^{+}e^{-} \to \left\{
		\begin{array}{c}
		\nu \bar{\nu}H \;\; (W^{+}W^{-}\hbox{ fusion})  \\
		e^{+}e^{-}H \;\; (Z^{0}Z^{0}\hbox{ fusion})
	\end{array} \right. \;\; .
	\label{eq:gbfuse}
\end{equation}
The capabilities of linear colliders for the Higgs-boson search (and 
for a rich variety of other investigations) have been summarized 
by Murayama and Peskin \cite{murapesk} and in the report of the 
ECFA/DESY Linear Collider Working Group \cite{zerwas}.  Thorough 
searches and incisive determinations of Higgs-boson couplings are 
possible.  It is plausible that, by measuring the $e^{+}e^{-} \to HZ$ 
excitation curve for a light Higgs boson, $M_{H}$ could be determined 
as well as $M_{W}$.  A typical estimate is
\begin{equation}
	\delta M_{H} \approx 60\mevcc 
	\sqrt{\frac{100\fb^{-1}}{\mathcal{L}}}\; , \hbox{ for }M_{H} = 
	100\gevcc \; .
	\label{eq:lcwid}
\end{equation}

By shining high-power, high-repetition-rate, eV-energy lasers onto 
the $e^{+}$ and $e^{-}$ (or $e^{-}$ and $e^{-}$) beams in a linear 
collider, it is possible to create a $\gamma\gamma$ collider with 
definite polarization and useful luminosity.  Such an instrument 
would be well suited to the study of the formation reaction
\begin{equation}
	\gamma \gamma \to H \to b\bar{b} \;\; ,
	\label{eq:ggtoH}
\end{equation}
with a rate proportional to $\Gamma(H\to\gamma\gamma)\Gamma(H\to 
b\bar{b})/\Gamma(H \to\hbox{all})$.  Knowing the $b\bar{b}$ branching 
ratio, one therefore has a direct determination of the 
$H\to\gamma\gamma$ coupling, an important diagnostic for the physics 
of electroweak symmetry breaking.

\subsubsection{A $\mu^{+}\mu^{-}$ Higgs Factory}
In common with the electron, the muon is an elementary lepton at our 
current limits of resolution.  Its energy is not shared among many 
partons, so the muon is a more efficient delivery vehicle for high 
energies than is the composite proton.  Because the muon is so massive, 
synchrotron radiation does not represent a barrier to small, 
high-energy, circular machines---as it does for electrons.

Beyond the suggestion of these practical advantages, muons offer a 
possibly decisive physics advantage.  The great seduction of a First 
Muon Collider is that the cross section for the reaction
$\mu^{+}\mu^{-} \rightarrow H$, 
direct-channel formation of the Higgs boson, is larger than the cross 
section for $e^{+}e^{-}\rightarrow H$ by a factor 
$(m_{\mu}/m_{e})^{2} \approx 42,750$.  This is a very large factor.  
The tantalizing question is whether it is large enough to make 
possible a ``Higgs factory'' with the luminosities that may be 
achieved in \mumu s.  In $e^{+}e^{-}$ collisions, as we have remarked, the 
$s$-channel formation cross section is hopelessly small.  That is why 
the associated-production reaction $e^{+}e^{-}\rightarrow HZ$ has 
become the preferred search mode at LEP~2.

The properties of the muon also raise challenges to the construction 
and exploitation of a \mumu.  The muon is not free: it doesn't come 
out of a bottle like the proton or boil off a metal plate like the 
electron.  On the other hand, it is readily produced in the decay 
$\pi \rightarrow \mu\nu$.  Still, gathering large numbers of muons in 
a dense beam is a formidable engineering challenge, and the focus of 
much of the R\&D effort over the next few years.  The muon is also 
not stable, but decays with a lifetime of 2.2~$\mu$s into $\mu^{-} 
\rightarrow e^{-}\bar{\nu}_{e}\nu_{\mu}$.  We must act fast to 
capture, cool, accelerate, and use muons, and must be able to 
replenish the supply quickly.  Multiply 2.2~$\mu$s by whatever 
Lorentz $(\gamma)$ factor you like for a muon collider, it is still a 
very short time.

The important possibility that a \mumu\ can operate as a Higgs factory 
has been studied extensively \cite{higgs,higgz}.  If the Higgs boson is light 
($M_{H}\ltap 2M_{W}$), and therefore narrow, then the muon's large 
mass makes it thinkable that the reactions
\begin{displaymath}
	\mu^{+}\mu^{-} \rightarrow H \rightarrow b\bar{b}\hbox{ and other 
	modes}
\end{displaymath}
will occur with a large rate that will enable a comprehensive study 
of the properties of the Higgs boson.  We assume that a light Higgs 
boson has been found, and that its mass has been determined with an 
uncertainty of $\pm(100\,\hbox{-}\,200)\mevcc$ \cite{gunion}.  Then 
suppose that an optimized machine is built with $\sqrt{s} = M_{H}$.

The muon's mass confers another important instrumental advantage: the 
momentum spread of a muon collider is naturally small, and can be made 
extraordinarily small.  The Higgs factory \cite{ank} would operate in two modes:
\begin{itemize}
	\item  modest luminosity ($0.05\fb^{-1}/\hbox{year}$) and high 
	momentum resolution ($\sigma_{p}/p = 3 \times 10^{-5}$);

	\item  standard luminosity ($0.6\fb^{-1}/\hbox{year}$) and normal momentum 
	resolution ($\sigma_{p}/p = 10^{-3}$).
\end{itemize}
At high resolution, the spread in c.m.\ energy is comparable to the 
natural width of a light Higgs boson: $\sigma_{\sqrt{s}} \approx 
\hbox{a few}\mev \approx \Gamma(H\rightarrow\hbox{ all})$.  At normal 
resolution, $\sigma_{\sqrt{s}} \gg \Gamma(H\rightarrow\hbox{ all})$.   It is 
worth remarking that the Higgs factory would be small, with a 
circumference of just 380 meters, and that the number of turns a muon 
makes in one lifetime is 820.

The first order of business is to run in the high-resolution mode to 
determine the Higgs-boson mass with exquisite precision.  The 
procedure contemplated is to scan a large number of points (determined 
by $2\Delta M_{H}/\sigma_{\sqrt{s}} \approx 100$), each with enough 
integrated luminosity to establish a three-standard-deviation 
excess.  If each point requires an integrated luminosity  
$0.0015\fb^{-1}$, then the scan requires $100 \times 0.0015\fb^{-1} = 
0.15\fb^{-1}$, about three nominal years of running.  The reward is 
that, after the scan, the Higgs-boson mass will be known with an 
uncertainty of $\Delta M_{H} \approx \sigma_{\sqrt{s}} 
\approx 2\mevcc$, which is quite stunning.

Extended running in the form of a three-point scan of the Higgs-boson 
line at $\sqrt{s} = M_{H}, M_{H}\pm \sigma_{\sqrt{s}}$ would then 
make possible an unparalleled exploration of Higgs-boson properties.  
With an integrated luminosity of $0.4\fb^{-1}$ one may contemplate 
precisions of $\Delta M_{H}  \approx  0.1\mevcc$, 
$\Delta \Gamma_{H}  \approx  0.5\mev \approx \cfrac{1}{6}\Gamma_{H}$,
$\Delta(\sigma\cdot B(H\rightarrow b\bar{b}))  \approx  3\%$, and
$\Delta(\sigma\cdot B(H\rightarrow WW^{\star}))  \approx 15\%$. 

These are impressive measurements indeed.  The width of the putative 
Higgs boson is an important discriminant for supersymmetry, for it can 
range from the standard-model value to considerably larger values.  
Within the minimal supersymmetric extension of the standard model 
(MSSM), the ratio of the $b\bar{b}$ and $WW^{\star}$ yields is 
essentially determined by $M_{A}$, the mass of the \textsl{CP}-odd Higgs 
boson.  In the decoupling limit, $M_{A}\rightarrow \infty$, the MSSM 
reproduces the standard-model ratio.  Deviations indicate that $A$ is 
light.  In the most optimistic scenario, this measurement could 
determine $M_{A}$ well enough to guide the development of a 
second (\textsl{CP}-odd) Higgs factory using the reaction 
$\mu^{+}\mu^{-}\rightarrow A$.

Again, these remarkable measurements exact a high price.  At a 
luminosity of $0.05\fb^{-1}/\hbox{year}$, it takes 8 years to 
accumulate $0.40\fb^{-1}$ \textit{after the scan} to determine $M_{H}$ 
within machine resolution.  It is plain that this program becomes 
considerably more compelling if the Higgs-factory luminosity can be 
raised by a factor of 2 or 3---or more!  Let us note finally that the 
flux of decay electrons challenges the operation of silicon detectors 
close to the interaction point \cite{sili}.

\subsubsection{$\bar{p}p$ Collisions at the Tevatron}
\begin{figure}[t!]
	\centerline{\BoxedEPSF{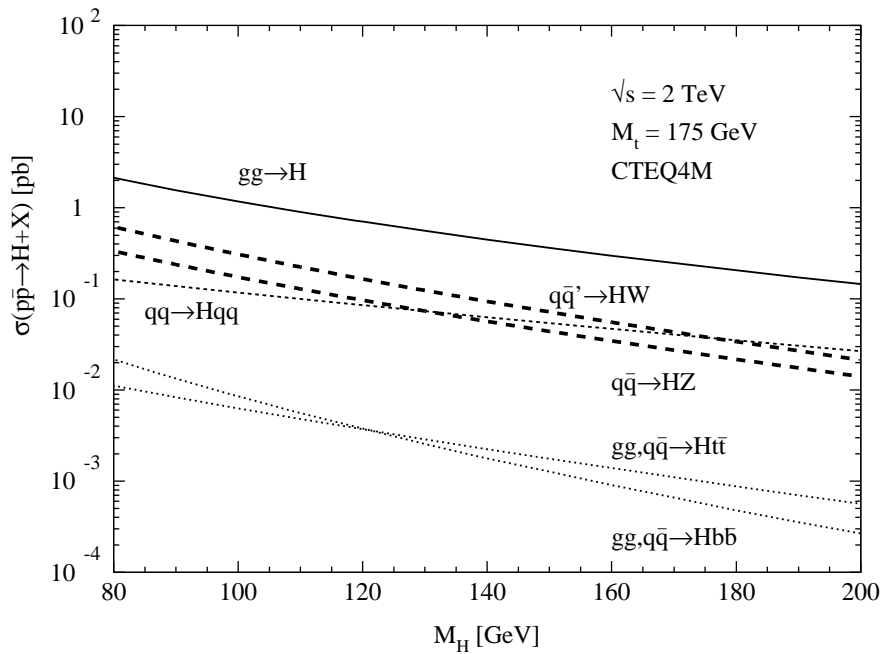  scaled 700}}
	\vspace*{6pt}
	\caption{Cross sections for Higgs-boson production in 2-TeV $\bar{p}p$ 
	collisions, from Ref.\ {\protect{\cite{spira}}}.}
	\protect\label{fig:Hsig}
\end{figure}

The cross sections for Higgs-boson production at the Tevatron are 
shown in Figure \ref{fig:Hsig} \cite{spira}.  The values---no larger than a 
few picobarns---highlight the need for large integrated luminosity 
and favorable branching fractions.  At the same time, many processes 
become accessible once the integrated luminosity exceeds  
$\hbox{a few}\fb^{-1}$.

The most promising channel for searches at the Tevatron will be the 
$b\bar{b}$ mode, for which the branching fraction exceeds about 
50\% throughout the region preferred by supersymmetry and the 
precision electroweak data.  
At the Tevatron, the direct production of a light Higgs boson in 
gluon-gluon fusion $gg \rightarrow H \rightarrow b\bar{b}$ is swamped by 
the ordinary QCD production of $b\bar{b}$ pairs. Even with an 
integrated luminosity of $30\fb^{-1}$, the experiments anticipate 
only $<1\hbox{-}\sigma$ excess, with plausible invariant-mass 
resolution.  It will be possible to calibrate the $b\bar{b}$ mass 
resolution over the region of the Higgs search in Run II, which aims 
to accumulate $2\fb^{-1}$: the 
electroweak production of $Z^{0}\rightarrow b\bar{b}$ should stand 
well above background and be clearly observable in Run II \cite{cdfZ}.

The high background in the $b\bar{b}$ channel means that special 
topologies must be employed to improve the ratio of signal to background 
and the significance of an observation.  The high luminosities that can 
be contemplated for a future run argue that the associated-production reactions
    \bothdk{\bar{p}p}{H}{W+\hbox{anything}}{\ell\nu}{b\bar{b}}{HWchan} 
and
	\bothdk{\bar{p}p}{H}{Z+\hbox{anything}}{\ell^{+}\ell^{-}+\nu\bar{\nu}}{b\bar{b}}{HZchan}
are plausible candidates for a Higgs discovery at the Tevatron 
\cite{marsw}.  

The prospects for exploiting these topologies were explored in detail 
in connection with the Run II Supersymmetry / Higgs Workshop at 
Fermilab
\cite{SHW}.  Taking into account what is known, and what might 
conservatively be expected, about sensitivity, mass resolutions, and 
background rejection, these investigations show that it is unlikely 
that a standard-model Higgs boson could be observed in Tevatron Run 
II.  (Note, however, that the ability to use $W\rightarrow q\bar{q}$ 
decays would markedly increase the sensitivity.)  
The prospects are much brighter for Run III.  Indeed, the sensitivity 
to a light Higgs boson is what motivates the integrated luminosity of 
$30\fb^{-1}$ specified for Run III.  

\begin{figure}[t!]
	\centerline{\BoxedEPSF{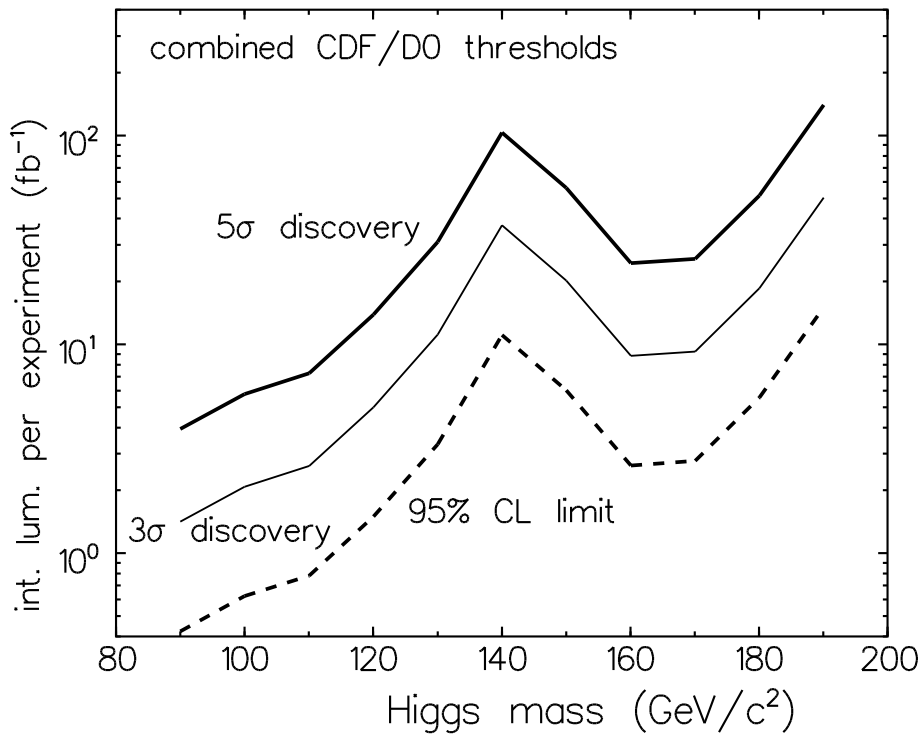  scaled 700}}
	\vspace*{6pt}
	\caption{Integrated luminosity projected for the 
	detection of a standard-model Higgs boson at the Tevatron Collider.}
	\protect\label{fig:Hportee}
\end{figure}

The detection strategy evolved in the Supersymmetry / Higgs Workshop 
involves combining the $HZ$ and $HW$ signatures of \eqn{eq:HWchan} and 
\eqn{eq:HZchan}, 
and adding the data from the CDF and D\O\ detectors.  Prospects are 
summarized in Figure \ref{fig:Hportee}, which shows as a function of 
the Higgs-boson mass the luminosity required for exclusion at 95\%\ 
confidence level (dashed line), three-standard-deviation evidence 
(thin solid line), and five-standard-deviation discovery (thick solid 
line).  We see that an integrated luminosity of $2\fb^{-1}$, expected 
in Run II, is insufficient for a convincing observation of a 
standard-model Higgs boson with a mass too large to be observed at 
LEP~2.  However, a 95\%\ CL exclusion is possible up to about 
$125\gevcc$.  On the other hand, about $10\fb^{-1}$ would permit 
detailed study of a standard-model Higgs boson discovered at LEP~2.  
If the Higgs boson lies beyond the reach of LEP~2, $M_{H} \gtap (100 
\hbox{ -- }105)\gevcc$, then a 5-$\sigma$ discovery will be possible in 
a future Run III of the Tevatron ($30\fb^{-1}$) for masses up to 
about $(125\hbox{ -- }130)\gevcc$.  This prospect is the most powerful 
incentive we have for Run III.  Over the range of masses accessible in 
associated production at the Tevatron, it should be possible to determine the mass of the 
Higgs boson to $\pm (1\hbox{ -- }3)\gevcc$.

Recent studies \cite{hantz} suggest that it may be possible to extend 
the reach of the Tevatron significantly by making use of the 
real-$W$--virtual-$W$ ($WW^{*}$) decay modes for Higgs boson produced 
in the elementary reaction $gg \to H$.  As we saw in the discussion 
leading up to Figure \ref{fig:HHdk}, the $WW^{*}$ channel has the largest 
branching fraction for $M_{H} \gtap 140\gevcc$.  According to the 
analysis summarized in Figure \ref{fig:Hportee}, the large cross 
section $\times$ branching fraction of the $gg \to H \to WW^{*}$ mode 
extends the 3-$\sigma$ detection sensitivity of Run III into the 
region $145\gevcc \ltap M_{H} \ltap 180\gevcc$.  This is an extremely 
exciting opportunity, and it is important that the $WW^{*}$ proposal 
receive independent critical analysis.  For the moment, it appears 
that the determination of the Higgs-boson mass would have limited 
precision, perhaps $\delta M_{H} \approx 30\gevcc$ \cite{taohan}.  
This question also requires additional study.

\subsubsection{$pp$ Collisions at the LHC}
Many significant advances have informed preparations for experiments 
at the Large Hadron Collider.  These include new or enhanced detector 
components and improved integration of individual elements into a 
high-performance detector, refined Monte Carlo tools, the evolution of 
new techniques for computing multiparton amplitudes, and progress in 
accelerator technology.  The capabilities of the LHC experiments to 
search for, and study, the Higgs boson are thoroughly documented in 
the Technical Proposals \cite{tdrs}.  I will confine myself here to a 
few summary comments.

A 5-$\sigma$ discovery is possible up to $M_{H}\approx 
	800\gevcc$ in a combination of the channels
	\bothdk{H}{Z}{\;Z}{\ell^{+}\ell^{-}}{\ell^{+}\ell^{-},}{HZ4l}
	\twodk{H}{\:W}{\ell\nu}{b\bar{b}}{HWlb}
	and 
\begin{displaymath}
	H \rightarrow \gamma\gamma \hbox{ or perhaps }\tau^{+}\tau^{-}.
\end{displaymath}
The reach of LHC experiments can be extended by making use of the 
channels 
	\bothdk{H}{Z}{\;Z}{\ell^{+}\ell^{-}\hbox{ or }\nu\bar{\nu}}{\hbox{jet 
	jet},}{HZlnu}
	and
	\bothdk{H}{W}{\:W}{\ell\nu}{\hbox{jet jet}.}{HZjj}
For Higgs-boson masses below about $300\gevcc$, it should be possible 
to determine the Higgs mass to 100-$300\mevcc$ \cite{gunion}.  For a 
recent exposition of the prospects for Higgs-boson searches from LEP 
to the LHC, see the lectures by Dittmar \cite{ditt}.

\section{Higgs Physics beyond the Standard Model}
In this final lecture, I want to review some indications for physics 
beyond the standard model, and explore some possibilities for the new 
phenomena we might encounter.  To begin, I want to revisit a 
longstanding, but usually unspoken, challenge to the completeness of 
the electroweak theory as we have defined it: the vacuum energy 
problem.  I do so not only for its intrinsic interest, but also to 
raise the question, ``Which problems of completeness and 
consistency do we worry about at a given moment?''  It is perfectly 
acceptable science---indeed, it is often essential---to put certain 
problems aside, in the expectation that we will return to them at the 
right moment.  What is important is never to forget that the problems 
are there, even if we do not allow them to paralyze us.  Then I will 
return to the significance of the 1-TeV scale, and move on to brief 
comments on supersymmetry and technicolor.  The final topic of this 
lecture is the problem of fermion masses, which is undoubtedly linked 
to the question of electroweak symmetry breaking, but calls for new 
insights that will go beyond the standard model.
\subsection{The Vacuum Energy Problem}
For our simple choice \eqn{SSBpot} of the Higgs potential, the value of 
the potential at the minimum is
\begin{equation}
    V(\vev{\phi^{\dagger}\phi}) = \frac{\mu^{2}v^{2}}{4} = 
    - \frac{\abs{\lambda}v^{4}}{4} < 0.
    \label{minpot}
\end{equation}
Identifying $M_{H}^{2} = -2\mu^{2}$, we see that the Higgs potential 
contributes a field-independent constant term,
\begin{equation}
    \varrho_{H} \equiv \frac{M_{H}^{2}v^{2}}{8}.
    \label{eq:rhoH}
\end{equation}
I have chosen the notation $\varrho_{H}$ because the constant term in the 
Lagrangian plays the role of a vacuum energy density.  When we 
consider gravitation, adding a vacuum energy density 
$\varrho_{\mathrm{vac}}$ is equivalent to adding a cosmological constant 
term to Einstein's equation.  Although recent observations 
\cite{cosconst} raise the intriguing possibility that the cosmological 
constant may be different from zero, the essential observational fact 
is that the vacuum energy density must be very tiny indeed \cite{TdA},
\begin{equation}
    \varrho_{\mathrm{vac}} \ltap 10^{-46}\gev^{4}\; .
    \label{eq:rhovaclim}
\end{equation}
Therein lies the puzzle: if we take
$v = (G_F\sqrt{2})^{-\frac{1}{2}}  \approx 246\gev$ from 
\eqn{eq:vevdef} and insert the current experimental lower bound 
\cite{Hsearch}
$M_{H} \gtap 95\gevcc$ into \eqn{eq:rhoH}, we find that the 
contribution of the Higgs field to the vacuum energy density is
\begin{equation}
    \varrho_{H} \gtap 7.6 \times 10^{7}\gev^{4},
    \label{eq:rhoHval}
\end{equation}
some 54 orders of magnitude larger than the upper bound inferred from 
the cosmological constant.

What are we to make of this mismatch, which has been apparent 
\cite{Veltman:1975au,Linde:1974at,Dreitlein:1974,Weinberg:1989cp} for 
nearly a quarter of a century?  The fact that $\varrho_{H} \gg 
\varrho_{\mathrm{vac}}$ means that the smallness of the cosmological 
constant needs to be explained.  In a unified theory of the strong, 
weak, and electromagnetic interactions, other (heavy!) Higgs fields 
have nonzero vacuum expectation values that may give rise to still 
greater mismatches.  At a fundamental level, we can therefore conclude 
that a spontaneously broken gauge theory of the strong, weak, and 
electromagnetic interactions---or merely of the electroweak 
interactions---cannot be complete.  Either we must find a separate 
principle that zeroes the vacuum energy density of the Higgs field, or 
we may suppose that a proper quantum theory of gravity, in combination 
with the other interactions, will resolve the puzzle of the 
cosmological constant.  In an interesting paper that prefigured the 
idea of ``large'' extra dimensions, van der Bij \cite{vdBij} has 
argued that because gravity and the Higgs field are both universal, 
they must be linked, perhaps in a spontaneously broken gravity in 
which the standard-model Higgs boson is the origin of the Planck mass.

The vacuum energy problem must be an important clue.  But to what?
\subsection{Why is the Electroweak Scale Small?}
In the first two lectures, we have outlined the electroweak theory, 
emphasized that the need for a Higgs boson (or substitute) is quite 
general, and reviewed the properties of the standard-model Higgs 
boson.  By considering a thought experiment, gauge-boson scattering 
at very high energies, we found a first signal for the importance of 
the 1-TeV scale.  Now, let us explore another path to the 1-TeV scale.

The $SU(2)_L \otimes U(1)_Y$ electroweak theory does not explain how the 
scale of electroweak symmetry breaking is maintained in the presence 
of quantum corrections.  The problem of the scalar sector can be 
summarized neatly as follows \cite{10}.  The Higgs potential is
\begin{equation}
      V(\phi^\dagger \phi) = \mu^2(\phi^\dagger \phi) +
\abs{\lambda}(\phi^\dagger \phi)^2 \;.
\end{equation}
With $\mu^2$ chosen to be less than zero, the electroweak symmetry is 
spontaneously broken down to the $U(1)$ of electromagnetism, as the 
scalar field acquires a vacuum expectation value that is fixed by the low-energy
phenomenology, 
\begin{equation}
	\vev{\phi} = \sqrt{-\mu^2/2|\lambda|} \equiv (G_F\sqrt 8)^{-1/2}
		\approx 175 {\rm \;GeV}\;.
		\label{hvev}
\end{equation}

Beyond the classical approximation, scalar mass parameters receive 
quantum corrections from loops that contain particles of spins 
$J=1, 1/2$, and $0$:
\begin{equation}
\BoxedEPSF{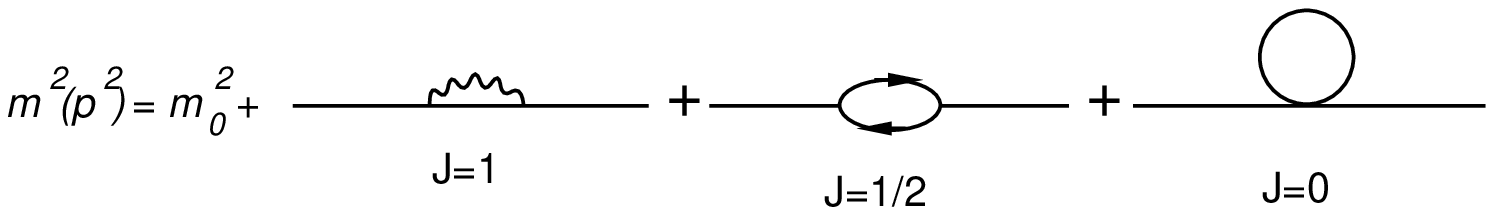  scaled 800}
\label{loup}
\end{equation}
The loop integrals are potentially divergent.  Symbolically, we may 
summarize the content of \eqn{loup} as
\begin{equation}
	m^2(p^2) = m^2(\Lambda^2) + Cg^2\int^{\Lambda^2}_{p^2}dk^2 
	+ \cdots \;,
	\label{longint}
\end{equation}
where $\Lambda$ defines a reference scale at which the value of 
$m^2$ is known, $g$ is the coupling constant of the theory, and the 
coefficient $C$ is calculable in any particular theory.  
Instead of dealing with the relationship between observables and 
parameters of the Lagrangian, we choose to describe the variation of 
an observable with the momentum scale.  In order for the mass shifts 
induced by radiative corrections to remain under control (\ie , not to 
greatly exceed the value measured on the laboratory scale), either 
$\Lambda$ must be small, so the range of integration is not 
enormous, or new physics must intervene to cut off the integral.

If the fundamental interactions are described by an 
$SU(3)_c\otimes SU(2)_L\otimes U(1)_Y$ gauge symmetry, \ie, by quantum
chromodynamics and the electroweak theory, then the 
natural reference scale is the Planck mass,
\begin{equation}
	\Lambda \sim M_{\rm Planck}  = 
	\left(\frac{\hbar c}{G_{\mathrm{Newton}}}\right)^{1/2} \approx 1.22 
	\times 10^{19} {\rm \; GeV}\;.
\end{equation}
In a unified theory of the strong, weak, and electromagnetic 
interactions, the natural scale is the unification scale,
\begin{equation}
	\Lambda \sim M_U \approx 10^{15}\hbox{-}10^{16} {\rm \; GeV}\;.
\end{equation}
Both estimates are very large compared to the scale of electroweak 
symmetry breaking \eqn{hvev}.  We are therefore assured that new physics must 
intervene at an energy of approximately 1~TeV, in order that the 
shifts in $m^2$ not be much larger than \eqn{hvev}.

Only a few distinct scenarios for controlling the 
contribution of the integral in \eqn{longint} can be envisaged.  The 
supersymmetric solution is especially elegant.  Exploiting the fact 
that fermion loops contribute with an overall minus sign (because of 
Fermi statistics), supersymmetry balances the contributions of fermion 
and boson loops.  In the limit of unbroken supersymmetry, in which the 
masses of bosons are degenerate with those of their fermion 
counterparts, the cancellation is exact:
\begin{equation}
	\sum_{{i={\rm fermions \atop + bosons}}}C_i\int dk^2 = 0\;.
\end{equation}
If the supersymmetry is broken (as it must be in our world), the 
contribution of the integrals may still be acceptably small if the 
fermion-boson mass splittings $\Delta M$ are not too large.  The 
condition that $g^2\Delta M^2$ be ``small enough'' leads to the 
requirement that superpartner masses be less than about 
$1\tevcc$.

A second solution to the problem of the enormous range of integration in 
\eqn{longint} is offered by theories of dynamical symmetry breaking such as 
technicolor. In technicolor models, the Higgs boson is composite, and 
new physics arises on the scale of its binding, $\Lambda_{\mathrm{TC}} \simeq 
O(1~{\rm TeV})$. Thus the effective range of integration is cut off, and 
mass shifts are under control.

A third possibility is that the gauge sector becomes strongly 
interacting. This would give rise to $WW$ resonances, multiple 
production of gauge bosons, and other new phenomena at energies of 1 TeV 
or so.  It is likely that a scalar bound state---a quasi-Higgs 
boson---would emerge with a mass less than about $1\tevcc$ \cite{Chanowitz:1998wi}.

We cannot avoid the conclusion that some new physics must occur on 
the \onetev.

\subsection{Supersymmetry}
The search for supersymmetry was discussed extensively here in Sierra 
Nevada by Daniel Treille \cite{treille} and Daniel Denegri 
\cite{denegri}, so I will restrict myself to a 
few general remarks about the motivation for supersymmetry on the 
electroweak scale, and its connection with string theory 
\cite{lykken,SteveM,Sallyd}.

\begin{figure}[t!]
	\centerline{\BoxedEPSF{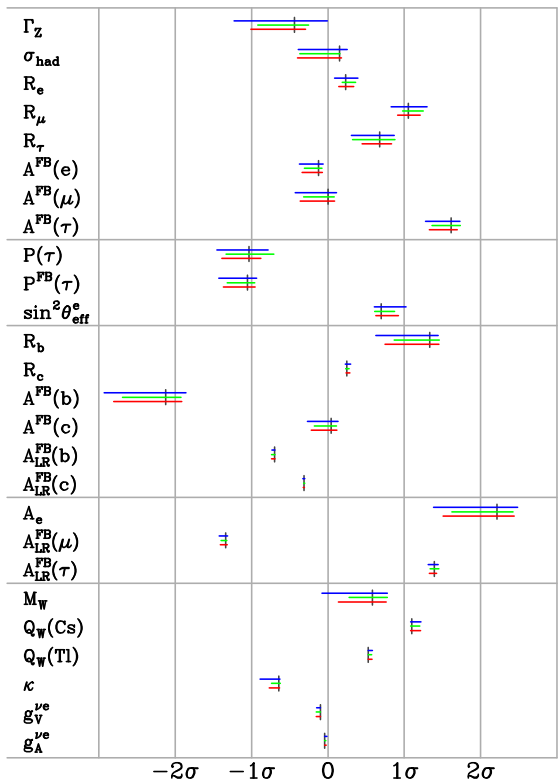  scaled 600}}
	\vspace*{6pt}
\caption{The range of best fit predictions of precision observables in
the supergravity model (upper horizontal lines), the $\mathbf{5} \oplus \mathbf{5^{*}}$
gauge-mediated model (middle lines), the 
$\mathbf{10} \oplus \mathbf{10^{*}}$ gauge-mediated model
(lower lines), and in the standard model at its global best fit value
(vertical lines), in units of standard deviation, from Ref.\ {\protect \cite{Erler:1998ur}}.}
	\protect\label{fig:allobs}
\end{figure}

One of the best phenomenological motivations for supersymmetry on the 
\onetev\ is that the minimal supersymmetric extension of the standard 
model so closely approximates the standard model itself.  A nice 
illustration of the small differences between predictions of 
supersymmetric models and the standard model is the compilation of 
pulls prepared by Erler and Pierce \cite{Erler:1998ur}, which is shown 
in Figure \ref{fig:allobs}.  This is a 
nontrivial property of new physics beyond the standard model, and a 
requirement urged on us by the unbroken quantitative success of the 
established theory.  On the aesthetic---or theoretical---side, supersymmetry 
is the maximal---indeed, unique---extension of Poincar\'{e} invariance.  
It also offers a path to the incorporation of gravity, since local 
supersymmetry leads directly to supergravity.  As a practical matter, 
supersymmetry on the \onetev\ offers a solution to the naturalness 
problem, and allows a fundamental scalar to exist at low energies.

When we combine supersymmetry with unification of the fundamental 
forces, we obtain a satisfactory prediction for the weak mixing 
parameter, $\sin^{2}\theta_{W}$, and a simple picture of 
coupling-constant unification \cite{deBoer:1996hd}.  Adding an assumption of universality, 
we are led naturally to a picture in which the top mass is linked 
with the electroweak scale, so that $m_{t} \approx v/\sqrt{2}$.  
Finally, the assumption of $R$-parity leads to a stable lightest 
supersymmetric particle, which is a natural candidate for the dark 
matter of the Universe.

Supersymmetry doubles the spectrum of fundamental particles.  We know 
that supersymmetry must be significantly broken in Nature, because the 
electron is manifestly not degenerate in mass with its scalar partner, 
the selectron.  It is interesting to contemplate just how different 
the world would have been if the selectron, not the electron, were 
the lightest charged particle and therefore the stable basis of 
everyday matter \cite{18params}.  If atoms were selectronic, there would be 
no Pauli principle to dictate the integrity of molecules.  As 
Dyson~\cite{dyson} and Lieb~\cite{lieb} demonstrated, transforming 
electrons and nucleons from fermions to bosons would cause all 
molecules to shrink into an insatiable undifferentiated blob.  
Luckily, there is no analogue of chiral symmetry to 
guarantee naturally small squark and slepton masses.  So while 
supersymmetry menaces us with an amorphous death, it is 
likely that a full understanding of supersymmetry will enable us to 
explain why we live in a universe ruled by the exclusion principle.

Many theorists take a step beyond supersymmetry to string theory, the 
only known consistent theory of quantum gravity \cite{quevedo,joepol}.  String theory 
aspires to unite all the fundamental interactions in one (and only 
one?) theory with few free parameters.  If successful, this program 
might explain the standard-model gauge group, unified extensions to the
\sm\ gauge symmetry, and the fermion content of the standard model. 
The defining ambition of string theory is to reconcile quantum 
mechanics and the implications of the uncertainty principle with 
general relativity and its guiding notion of a smooth spacetime. 

String theory makes several generic predictions for physics beyond the 
standard model: additional $U(1)$ subgroups of the unifying group lead 
to new gauge bosons, and additional colored fermions augment the 
spectrum of fundamental constituents.  It also requires general 
relativity and, for consistency, extra spacetime dimensions, some of 
which might be detectably large 
\cite{Antoniadis:1990ew,Lykken:1996fj,Arkani-Hamed:1998rs,Dienes:1998vh}. 
 And string theory requires supersymmetry, though not necessarily on 
the 1-TeV scale.

In spite of what doubters often say, there \emph{is} experimental support 
for string theory from accelerator experiments.  Superstrings predicted 
gravity in 1974 \cite{joel}, and LEP accelerator physicists detected tidal 
forces in 1993 \cite{tides}.  What more empirical evidence could one 
demand?

In a supersymmetric theory, two Higgs doublets are required to give 
masses to fermions with weak isospin $I_{3}=\cfrac{1}{2}$ and 
$I_{3}=-\cfrac{1}{2}$. Let us designate the two doublets as 
$\Phi_{1}$ and $\Phi_{2}$.  Before supersymmetry is broken, the scalar 
potential has the form
\begin{equation}
	V = \mu^{2}(\Phi_{1}^{2} + \Phi_{2}^{2}) + 
	\frac{g^{2}+g^{\prime2}}{8}(\Phi_{1}^{2} + \Phi_{2}^{2})^{2} + 
	\frac{g^{2}}{2}\left|\Phi_{1}^{*} \cdot \Phi_{2}\right|^{2} \; .
	\label{eq:susyhiggspot}
\end{equation}
By adding all possible soft supersymmetry-breaking terms, we raise the 
possibility that the electroweak symmetry will be broken.  We choose
\begin{equation}
	\begin{array}{c}
		\vev{\Phi_{1}} = v_{1} > 0 \; ,  \\[6pt]
		\vev{\Phi_{2}} = v_{2} > 0 \; ,
	\end{array}
	\label{eq:susyvevs}
\end{equation}
with $v_{1}^{2} + v_{2}^{2} = v^{2}$ and
\begin{equation}
	\frac{v_{2}}{v_{1}} \equiv \tan\beta \; .
	\label{eq:tanb}
\end{equation}

After the $W^{\pm}$ and $Z^{0}$ acquire masses, five spin-zero 
degrees of freedom remain as massive spin-zero particles: the 
lightest scalar $h^{0}$, a heavier neutral scalar $H^{0}$, two charged 
scalars $H^{\pm}$, and a neutral pseudoscalar $A^{0}$.  At tree 
level, we may express all the (pseudo)scalar masses in terms of 
$M_{A}$ and $\tan\beta$, to find
\begin{equation}
	M_{h^{0},H^{0}}^{2} = \frac{1}{2} \left\{ M_{A}^{2} + M_{Z}^{2} \mp \left[ 
	(M_{A}^{2} + M_{Z}^{2})^{2} - 4M_{A}^{2} M_{Z}^{2} \cos^{2}2\beta
	\right]^{1/2}\right\} \; ,
	\label{eq:hHmass}
\end{equation}
and
\begin{equation}
	M_{H^{\pm}}^{2} = M_{W}^{2} + M_{A}^{2} \; .
	\label{eq:chHmass}
\end{equation}
At tree level, there is a simple mass hierarchy, given by
\begin{eqnarray}
	M_{h^{0}} & < & M_{Z}|\cos2\beta|
	\nonumber  \\
	M_{H^{0}} & > & M_{Z}
	\label{eq:susymasshier}  \\
	M_{H^{\pm}} & > & M_{W}
	\nonumber \; ,
\end{eqnarray}
but there are very important \textit{positive} loop corrections to $M_{h^{0}}^{2}$ 
(proportional to $G_{F}m_{t}^{4}$) that were overlooked in the 
earliest calculations.  These loop corrections change the mass 
predictions very significantly.

\begin{figure}
	\centerline{\BoxedEPSF{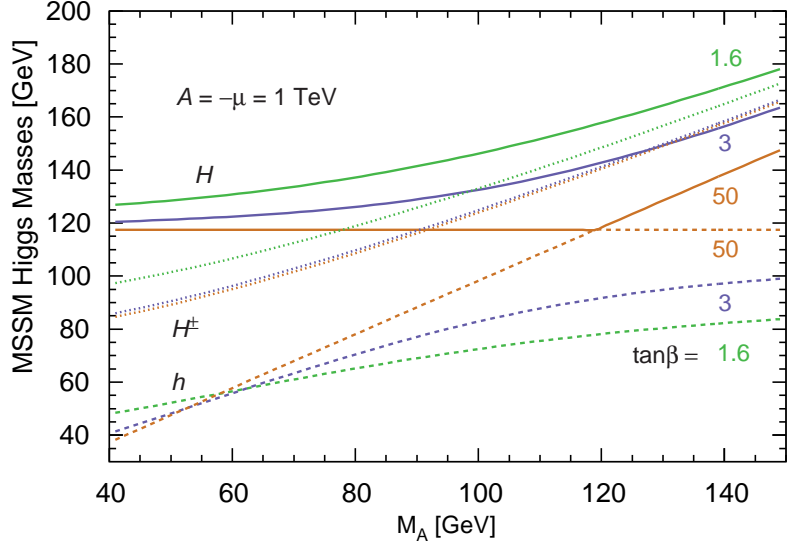  scaled 600}}
	\vspace*{6pt}
	\caption{Higgs-boson masses in the minimal supersymmetric standard 
	model as a function of $M_{A}$ for the case $A = -\mu = M_{S} = 
	1\tev$, for a top-quark mass $m_{t} = 175\gevcc$, from Ref. 
	{\protect \cite{Carena:1996bj}}.}
	\protect\label{fig:susyHmasses}
\end{figure}
The results of a full, modern calculation \cite{Carena:1996bj} are shown in 
Figure \ref{fig:susyHmasses}.  There we see that the mass of the 
lightest Higgs scalar is largest for large values of the pseudoscalar 
mass $M_{A}$ and in the limit of large $\tan\beta$.  

\begin{figure}[t!]
	\centerline{\BoxedEPSF{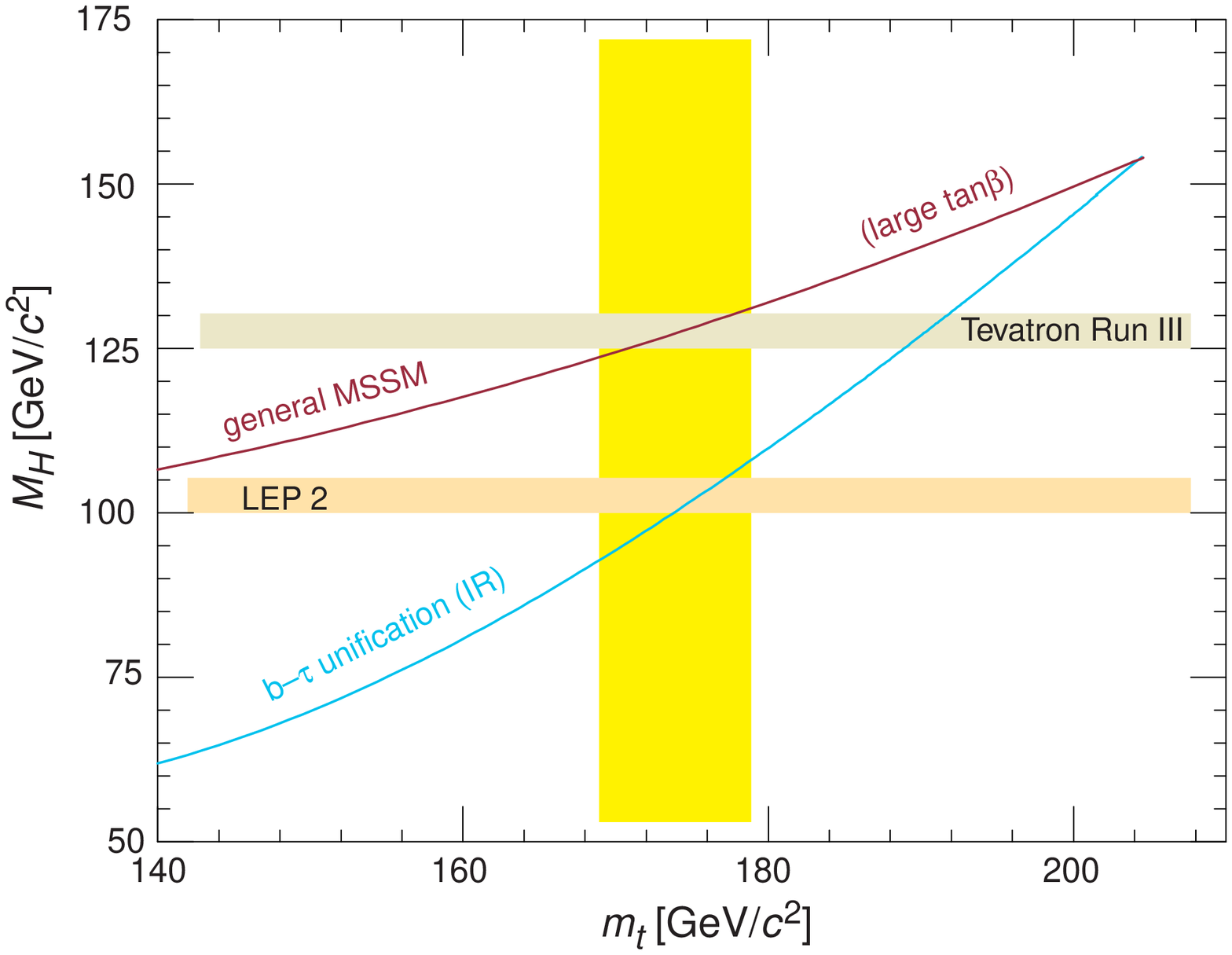  scaled 600}}
	\vspace*{6pt}
	\caption{Upper bounds on the mass of the lightest Higgs boson, as a 
	function of the top-quark mass, in two variants of the minimal 
	supersymmetric standard model.  The upper curve refers to a general 
	MSSM, in the large-$\tan\beta$ limit; the lower curve corresponds to 
	an infrared-fixed-point scenario with $b$-$\tau$ unification, from 
	Ref.\ {\protect \cite{Carena:1995bx}}.}
	\protect\label{fig:MHmt}
\end{figure}
Because the minimal supersymmetric standard model (MSSM) implies 
\textit{upper bounds} on the mass of the lightest scalar $h^{0}$, it 
sets attractive targets for experiment.  Two such upper bounds are 
shown as functions of the top-quark mass in Figure \ref{fig:MHmt}.  
The large-$\tan\beta$ limit of a general MSSM yields the upper curve; 
an infrared-fixed-point scheme with $b$-$\tau$ unification produces 
an upper bound characterized by the lower curve.  The vertical band 
shows the current information on $m_{t}$.  We see that the projected 
sensitivity of LEP~2 experiments covers the full range of 
lightest-Higgs masses that occur in the infrared-fixed-point scheme.  
The sensitivity promised by Run III of the Tevatron gives full 
coveralge of $h^{0}$ masses in the MSSM.  These are very intriguing 
experimental possibilities.  For further discussion, consult the LEP~2 
Yellow Book \cite{Carena:1996bj} and the Proceedings of the Tevatron 
Supersymmetry / Higgs Workshop \cite{SHW}.
\subsection{New Strong Dynamics}
Dynamical symmetry breaking offers a different solution to the 
naturalness problem of the electroweak theory: in technicolor, there 
are no elementary scalars.  We hope that solving the dynamics that 
binds elementary fermions into a composite Higgs boson and other $WW$ 
resonances will bring addition predictive power.  It is worth saying 
that technicolor is a far more ambitious program than global 
supersymmetry.  It doesn't merely seek to finesse the hierarchy 
problem, it aims to predict the mass of the Higgs surrogate.  Against 
the aesthetic appeal of supersymmetry we can weigh technicolor's excellent 
pedigree.  As we have seen in \S\ref{sub:cache},
the Higgs mechanism of the standard model 
is the relativistic generalization of the Ginzburg-Landau description 
of the superconducting phase transi\-tion.  Dynamical symmetry breaking 
schemes---technicolor and its relatives---are inspired by the
Bardeen--Cooper--Schrieffer theory of superconductivity, 
and seek to give a similar microscopic description of electroweak 
symmetry breaking.  

The dynamical-symmetry-breaking approach realized in technicolor theories
 is modeled upon our understanding of the superconducting phase 
transition \cite{varsovie,Chivukula:1996uy,18}. The macroscopic order parameter of the Ginzburg-Landau 
phenomenology corresponds to the wave function of superconducting 
charge carriers, which acquires a nonzero vacuum
expectation value in the 
superconducting state. The microscopic Bardeen-Cooper-Schrieffer 
theory \cite{19} identifies the dynamical origin of the order parameter with 
the formation of bound states of elementary fermions, the Cooper pairs of 
electrons. The basic idea of  technicolor is to replace the 
elementary Higgs boson with a fermion-antifermion bound 
state. By analogy with the superconducting phase transition, the dynamics 
of the fundamental technicolor gauge interactions among technifermions 
generate scalar bound states, and these play the role of the Higgs fields.

The elementary fermions---electrons---and 
gauge interactions---QED---needed to generate the scalar bound states are 
already present in the case of superconductivity. Could a scheme
 of similar economy account
for the transition that hides the electroweak symmetry?
Consider an $SU(3)_c\otimes SU(2)_L\otimes U(1)_Y$ theory of massless up and 
down quarks. Because the strong interaction is strong, and the electroweak 
interaction is feeble, we may treat the $SU(2)_L\otimes U(1)_Y$ 
interaction as a perturbation. For vanishing quark masses, QCD has an exact 
$SU(2)_L\otimes SU(2)_R$ chiral symmetry. At an energy scale 
$\sim\Lambda_{\mathrm{QCD}},$ the strong interactions become strong, fermion 
condensates appear, and the chiral symmetry is spontaneously broken
to the familiar flavor symmetry:
\begin{equation}
	SU(2)_L\otimes SU(2)_R \to SU(2)_V\;\; .
\end{equation}
 Three Goldstone bosons appear, one for 
each broken generator of the original chiral invariance. These were 
identified by Nambu~\cite{20} as three massless pions.

The broken generators are three axial currents whose couplings to pions are 
measured by the pion decay constant $f_\pi$. When we turn on the 
$SU(2)_L\otimes U(1)_Y$ electroweak interaction, the electroweak gauge 
bosons couple to the axial currents and acquire masses of order $\sim 
gf_\pi$. The mass-squared matrix,
\begin{equation}
	\mathcal{M}^{2} = \left(
		\begin{array}{cccc}
		g^{2} & 0 & 0 & 0  \\
		0 & g^{2} & 0 & 0  \\
		0 & 0 & g^{2} & gg^{\prime}  \\
		0 & 0 & gg^{\prime} & g^{\prime2}
	\end{array}
		 \right) \frac{f_{\pi}^{2}}{4} \; ,
	\label{eq:csbm2}
\end{equation}
(where the rows and columns correspond to $W^{+}$, $W^{-}$, $W_{3}$, 
and $\mathcal{A}$) has the same structure as the mass-squared matrix 
for gauge bosons in the standard electroweak theory.  Diagonalizing 
the matrix \eqn{eq:csbm2}, we find that $M_{W}^{2} = 
g^{2}f_{\pi}^{2}/4$ and $M_{Z}^{2} = 
(g^{2}+g^{\prime2})f_{\pi}^{2}/4$, so that 
\begin{equation}
	\frac{M_{Z}^{2}}{M_{W}^{2}} = \frac{(g^{2}+g^{\prime2})}{g^{2}} = 
	\frac{1}{\cos^{2}\theta_{W}}\; .
	\label{eq:wzrat}
\end{equation}
The photon emerges massless.

The massless pions thus disappear from the physical spectrum, 
having become the longitudinal components of the weak gauge bosons. 
Unfortunately, the mass acquired by the 
intermediate bosons is far smaller than required for a successful 
low-energy phenomenology; it is only~\cite{21} $M_W\approx 30\mevcc$.

The minimal technicolor model of Weinberg~\cite{22} and Susskind~\cite{23} 
transcribes the same ideas from QCD to a new setting.  The 
technicolor gauge group is taken to be $SU(N)_{\mathrm{TC}}$ (usually 
$SU(4)_{\mathrm{TC}}$), 
so the gauge interactions of the theory are generated by
\begin{equation}
	SU(4)_{\mathrm{TC}}\otimes SU(3)_c \otimes SU(2)_L \otimes U(1)_Y\; .
\end{equation}
The technifermions are a chiral doublet of massless color singlets
\begin{equation}
\begin{array}{cc}
	\left( \begin{array}{c} U \\ D \end{array} \right)_L & U_R, \;
D_R \; .
\end{array}
\end{equation}
With the electric charge assignments $Q(U)=\frac{1}{2}$ and
$Q(D)=-\frac{1}{2}$, the  
theory is free of electroweak anomalies. The ordinary fermions are all 
technicolor singlets. 

In analogy with our discussion of chiral symmetry breaking in QCD, we 
assume that the chiral TC symmetry is broken,
\begin{equation}
	SU(2)_L\otimes SU(2)_R\otimes U(1)_V\to SU(2)_V\otimes U(1)_V\; .
\end{equation}
Three would-be Goldstone bosons emerge. These are the technipions
\begin{equation}
\begin{array}{ccc}
\pi_T^+, & \pi_T^0, & \pi_T^-,
\end{array}
\end{equation}
for which we are free to {\em choose} the technipion decay constant as
\begin{equation}
	F_\pi = \left(G_F\sqrt{2}\right)^{-1/2} = 246\gev\; . \label{FPI}
\end{equation}
This amounts to choosing the scale on which technicolor becomes strong.
When the electroweak interactions are turned on, the technipions become the 
longitudinal components of the intermediate bosons, which acquire masses
\begin{equation}
\renewcommand\arraystretch{1.5}
\begin{array}{ccccc}
	M_W^2 & = & g^2F_\pi^2/4 & = & 
{\displaystyle \frac{\pi\alpha}{G_F\sqrt{2}\sin^2\theta_W}} \\
	M_Z^2 & = & \left(g^2+g^{\prime 2}\right)F_\pi^2/4 & = & 
M_W^2/\cos^2\theta_W
\end{array} \; ,
\renewcommand\arraystretch{1}
\end{equation}
that have the canonical Standard Model values, thanks to our choice 
(\ref{FPI}) 
of the technipion decay constant.

Technicolor shows how the generation of intermediate boson masses 
could arise without fundamental scalars or unnatural adjustments of 
parameters.  It thus provides an elegant solution to the naturalness 
problem of the Standard Model.  However, it has a major deficiency: it 
offers no explanation for the origin of quark and lepton masses, 
because no Yukawa couplings are generated between Higgs fields and 
quarks or leptons.  Consequently, technicolor serves as a reminder 
that there are two problems of mass: explaining the masses of the 
gauge bosons, which demands an understanding of electroweak symmetry 
breaking; and accounting for the quark and lepton masses, which 
requires not only an understanding of electroweak symmetry breaking 
but also a theory of the Yukawa couplings that set the scale of 
fermion masses in the standard model.  We can be confident that the 
origin of gauge-boson masses will be understood on the 1-TeV scale.  
We do not know where we will decode the pattern of the Yukawa 
couplings; I describe a possible approach in \S\ref{sub:fermas}.

To generate fermion mass, we may embed technicolor in a larger 
extended technicolor gauge group $G_{\mathrm{ETC}} \supset G_{\mathrm{TC}}$ that 
couples the quarks and leptons to technifermions \cite{etc}.  If the 
$G_{\mathrm{ETC}}$ gauge symmetry is broken down to $G_{\mathrm{TC}}$ 
at a scale $\Lambda_{\mathrm{ETC}}$, then the quarks and leptons can 
acquire masses
\begin{equation}
	m \sim \frac{g_{\mathrm{ETC}}^{2}F_{\pi}^{3}}{\Lambda_{\mathrm{ETC}}^{2}}\; .
	\label{eq:mETC}
\end{equation}
The generation of fermion mass is where all the experimental threats 
to technicolor arise.  The rich particle content of ETC models 
generically leads to quantum corrections that are in conflict with 
precision electroweak measurements \cite{lizsim}.  Moreover, if 
quantum chromodynamics is a good model for the chiral-symmetry 
breaking of technicolor, then extended technicolor produces 
flavor-changing neutral currents at uncomfortably large levels.  We 
conclude that QCD must not provide a good template for the technicolor 
interaction.

How could TC dynamics be different from QCD dynamics?  Most modern 
implementations of dynamical symmetry breaking invoke multiple scales 
\cite{24} to reconcile the generation of fermion masses with the 
constraints on flavor-changing neutral currents.  In traditional ETC, 
the TC interaction is precociously asympototically free.  If instead 
the technicolor interaction remains strong all the way from $F_{\pi}$ 
to $\Lambda_{\mathrm{ETC}}$, the link \eqn{eq:mETC} between 
$F_{\pi}$, $\Lambda_{\mathrm{ETC}}$, and fermion masses is modified to 
\begin{equation}
	m \sim \frac{g_{\mathrm{ETC}}^{2}F_{\pi}^{2}}{\Lambda_{\mathrm{ETC}}}\; ,
	\label{eq:walkfm}
\end{equation}
so that the scale $\Lambda_{\mathrm{ETC}}$ that produces the observed 
fermion masses can be much larger than before.  A high ETC scale in 
turn suppresses flavor-changing neutral-current processes to 
acceptable levels.

How could the $\beta$ function of the technicolor interaction be small 
over a broad range of energy scales?  To cancel the antiscreening 
contribution of gauge bosons, it is necessary to introduce either 
many fermions, or fermions in higher representations of the gauge 
group.  Adding more fermions enlarges the chiral symmetries and thus 
increases the number of (pseudo-)Goldstone bosons that arise.  The 
enriched spectrum of such a model could contain light resonances.  
Eichten, Lane, and Womersley \cite{elworm} have investigated a model 
in which the mesons built of techniquarks are relatively light.  The 
spectrum includes technipions $\pi_{\mathrm{T}}^{+}$, 
$\pi_{\mathrm{T}}^{0}$, and $\pi_{\mathrm{T}}^{-}$ with masses around 
$100\gevcc$, and technivector mesons $\rho_{\mathrm{T}}$ and 
$\omega_{\mathrm{T}}$ with masses around $250\gevcc$.  They have 
shown that new signatures,
\dkp{\omega_{T}}{\gamma}{\pi_{T}^{0}}{b\bar{b}}
and
\dkp{\rho_{T}^{+}}{W^{+}}{\pi_{T}^{0}}{b\bar{b}}
hold promise for incisive searches in Run II of the Tevatron Collider.

\subsection{The Problem of Fermion Masses {\protect{\label{sub:fermas}}}}
\begin{figure}
	\centerline{\BoxedEPSF{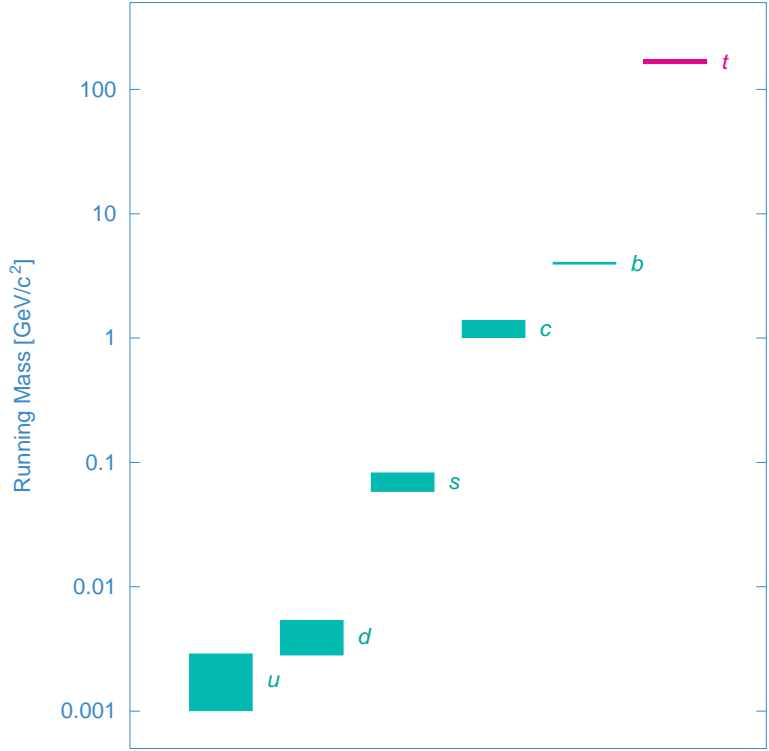  scaled 600}}
	\vspace*{6pt}
	\caption{Running masses [$m_{q}(m_{q})]$ of the quarks in the 
	$\overline{\mathrm{MS}}$ scheme.}
	\protect\label{fig:masses}
\end{figure}

The example of technicolor serves to emphasize that solving the origin 
of electroweak symmetry breaking will not necessarily give us insight 
into the origin of fermion masses.  How are we to make sense of the 
puzzling pattern of quark masses, for example, shown in Figure \ref{fig:masses}?
Like coupling constants, masses depend upon the momentum scale on 
which they are defined.  The values plotted in Figure 
\ref{fig:masses} are defined in the modified--minimal-subtraction 
($\overline{\mathrm{MS}}$) scheme, and evaluated at the mass of each 
quark (for $c,b,t$) or at $1\gev$ (for $u,d,s$).

The running of quark and lepton masses makes it evident that the pattern we 
perceive in our low-energy experiments is influenced not only by the 
underlying pattern, but also by the evolution from the scale on which 
masses are set down to our scale.  It is then no surprise that the 
pattern we discern seems irrational.  Perhaps an underlying order 
might show itself on some other scale.

It is helpful to examine this idea in a specific framework.  A simple 
and convenient example is provided by the $SU(5)$ unified theory of 
quarks and leptons, and of the strong, weak, and electromagnetic 
interactions.  In the $SU(5)$ theory, spontaneous symmetry breaking 
occurs in two steps.  First, a $\mathbf{24}$ of scalars breaks
\begin{equation}
	SU(5) \to SU(3)_{c}\otimes SU(2)_{L}\otimes U(1)_{Y} \; ,
	\label{su5b1}
\end{equation}
and gives extremely large masses to the leptoquark gauge bosons 
$X^{\pm 4/3}$ and $Y^{\pm 1/3}$.  The $\mathbf{24}$ does not occur in 
the $\bar{\mathsf{L}}\mathsf{R}$ products 
\begin{eqnarray}
	\mathbf{5}^{\ast}\otimes \mathbf{10} & = & \mathbf{5} \oplus \mathbf{45}
	\label{su5fm}  \\
	\mathbf{10} \otimes \mathbf{10} & = & \mathbf{5}^{\ast} \oplus 
	\mathbf{45}^{\ast} \oplus \mathbf{50}^{\ast}
	\nonumber
\end{eqnarray}
that generate fermion masses, so the quarks and leptons escape large 
masses at tree level.  In a second step, a $\mathbf{5}$ of scalars 
(which contains the standard-model Higgs doublet) breaks
\begin{equation}
	SU(3)_{c}\otimes SU(2)_{L}\otimes U(1)_{Y} \to SU(3)_{c}\otimes U(1)_{\mathrm{em}} \; ,
	\label{su5b2}
\end{equation}
and endows the fermions with mass.  This pattern of spontaneous 
symmetry breaking relates quark and lepton masses at the unification 
scale, predicting that
\begin{equation}
	\left.
	\begin{array}{c}
		m_{e} = m_{d}  \\
		m_{\mu} = m_{s}  \\
		m_{\tau} = m_{b}
	\end{array}
	\right\}\hbox{  at the unification scale }M_{U}.
	\label{eq:btauuni}
\end{equation}
The masses of the charge-$\cfrac{2}{3}$ quarks, $m_{u}$, $m_{c}$, 
$m_{t}$, are separate parameters.

Within the $SU(5)$ framework, it is straightforward to compute the 
evolution of the masses from $M_{U}$ to another scale $\mu$ \cite{Buras:1978yy}.  In 
leading logarithmic approximation, we find that

\begin{eqnarray}
	\ln\left[m_{u,c,t}(\mu)\right] & \approx & \ln\left[m_{u,c,t}(M_{U})\right]
	+ \frac{12}{33-2n_{f}}\ln\left(\frac{\alpha_{3}(\mu)}{\alpha_{U}}\right) 
	\label{eq:Uprun}\\
	 & & + 
	\frac{27}{88-8n_{f}}\ln\left(\frac{\alpha_{2}(\mu)}{\alpha_{U}}\right)
	- 
	\frac{3}{10n_{f}}\ln\left(\frac{\alpha_{1}(\mu)}{\alpha_{U}}\right) 
	\; , \nonumber
\end{eqnarray}
\begin{eqnarray}
	\ln\left[m_{d,s,b}(\mu)\right] & \approx & \ln\left[m_{d,s,b}(M_{U})\right]
	+ \frac{12}{33-2n_{f}}\ln\left(\frac{\alpha_{3}(\mu)}{\alpha_{U}}\right) 
	\label{eq:Downrun} \\
	 & & + 
	\frac{27}{88-8n_{f}}\ln\left(\frac{\alpha_{2}(\mu)}{\alpha_{U}}\right)
	+ 
	\frac{3}{20n_{f}}\ln\left(\frac{\alpha_{1}(\mu)}{\alpha_{U}}\right) 
	\; , \nonumber
\end{eqnarray}
and
\begin{eqnarray}
	\ln\left[m_{e,\mu,\tau}(\mu)\right] & \approx & \ln\left[m_{e,\mu,\tau}(M_{U})\right]
	 \label{eq:Leprun}\\
	 & & + 
	\frac{27}{88-8n_{f}}\ln\left(\frac{\alpha_{2}(\mu)}{\alpha_{U}}\right)
	- 
	\frac{27}{20n_{f}}\ln\left(\frac{\alpha_{1}(\mu)}{\alpha_{U}}\right) 
	\; , \nonumber
\end{eqnarray}
where $n_{f}$ is the number of quark or lepton flavors.
The running of the quark masses receives contributions from the 
color, weak-isospin, and weak-hypercharge interactions, whereas the 
color force does not influence the evolution of lepton masses.  
Accordingly, the quark masses run significantly, while the lepton 
masses run much less. [To 
simplify a bit, I have omitted the Higgs-boson contributions, which 
are important in the case of the heavy top quark.\footnote{A proper 
treatment would also take account of fermion (especially top!) 
threshold effects, and of the large $Ht\bar{t}$ couplings 
\cite{largeHY}.}]

\begin{figure}
	\centerline{\BoxedEPSF{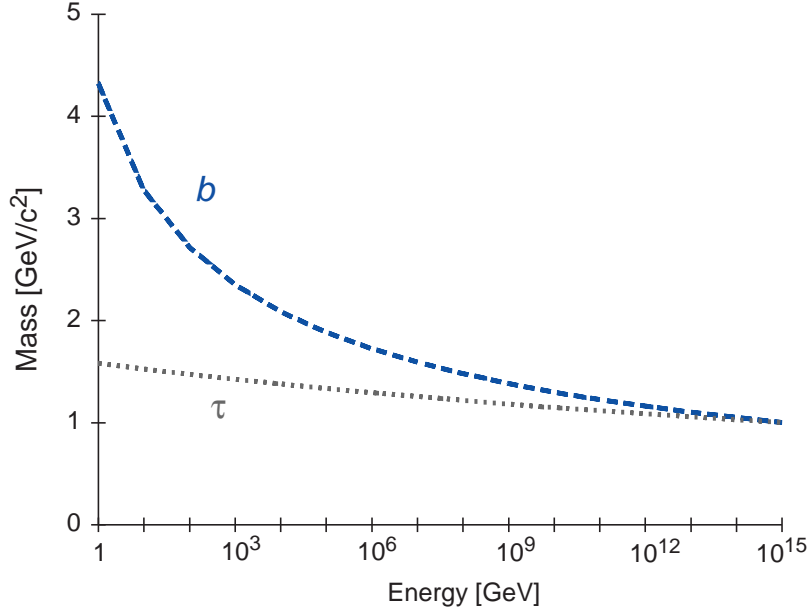  scaled 600}}
	\vspace*{6pt}
	\caption{Evolution of the $\tau$-lepton and $b$-quark masses from the 
	unification scale to low energies.}
	\protect\label{fig:runbt}
\end{figure}

The classic intriguing prediction of the $SU(5)$ theory involves the 
masses of the $b$ quark and the $\tau$ lepton, which are degenerate 
at the unification point.  To compare $m_{b}$ and $m_{\tau}$ on other 
scales, we combine \eqn{eq:Downrun} and \eqn{eq:Leprun} to write
\begin{eqnarray}
	\ln\left[\frac{m_{b}(\mu)}{m_{\tau}(\mu)}\right] & \approx & 
	\ln\left[\frac{m_{b}(M_{U})}{m_{\tau}(M_{U})}\right]
	 + \frac{12}{33-2n_{f}}\ln\left(\frac{\alpha_{3}(\mu)}{\alpha_{U}}\right) 
	\label{eq:btau}\\
	 & & \phantom{+ 
	\frac{27}{88-8n_{f}}\ln\left(\frac{\alpha_{2}(\mu)}{\alpha_{U}}\right)}
	- 
	\frac{3}{2n_{f}}\ln\left(\frac{\alpha_{1}(\mu)}{\alpha_{U}}\right) \; .
	\nonumber
\end{eqnarray}
The first term on the right-hand side vanishes, according to 
\eqn{eq:btauuni}.  If we consider the case $n_{f}=6$ with 
$1/\alpha_{U}=40$, $1/\alpha_{s}(\mu)=5$, and $1/\alpha_{1}(\mu)=65$, 
the evolution of $m_{b}$ and $m_{\tau}$ is shown in 
Figure \ref{fig:runbt}.  
At the low scale, we compute
\begin{equation}
	m_{b} = 2.91 m_{\tau} \approx 5.16\gevcc \; ,
	\label{eq:unitaub}
\end{equation}
in suggestive agreement with what we know from experiment.  The same 
procedure would lead to predictions for the first and second 
generations, at a scale $\mu \approx 1\gev$, namely
\begin{equation}
	\begin{array}{rcl}
		\underbrace{{\displaystyle \frac{m_{s}}{m_{d}}}} & = & 
		\underbrace{{\displaystyle \frac{m_{\mu}}{m_{e}}}}  \\
		\approx 20 &  & \approx 200
	\end{array}\; ,
	\label{eq:gen12}
\end{equation}
which is less successful.

A more elaborate symmetry-breaking scheme---for example, adding a 
$\mathbf{45}$ 
of scalars---can change the relationship between the electron mass and 
the down-quark mass at the unification scale.  A scheme that results in
\begin{equation}
	\begin{array}{ccc}
		m_{s} = \cfrac{1}{3}m_{\mu}, & m_{d} = 3 m_{e} & \hbox{at }M_{U}
	\end{array}
	\label{eq:scheme2}
\end{equation}
leads to the low-energy predictions
\begin{equation}
	\left.
	\begin{array}{c}
		m_{s} \approx \cfrac{4}{3}m_{\mu}  \\
		m_{d} \approx 12 m_{e}
	\end{array}
	\right\}\;\;\hbox{ at }\mu \approx 1\gev \; .
	\label{eq:modunitaub}
\end{equation}

In the 1990s, this kind of analysis has given rise to a new cottage 
industry for understanding the pattern of fermion masses---including 
neutrino masses.  Begin with a plausible unified theory, \eg, 
supersymmetric $SU(5)$, with its advantages for coupling-constant 
unification and the low-energy prediction for the weak mixing 
parameter $\sin^{2}\theta_{W}$, or supersymmetric $SO(10)$, to include 
massive neutrinos.  Then find ``textures,'' simple patterns of Yukawa 
matrices that lead to successful predictions for masses and mixing 
angles.  Interpret these in terms of patterns of symmetry breaking.  
Finally, seek a derivation---or at least some motivation---for the 
explicit entry.  I think this is a very interesting strategy; whether 
progress will be rapid or slow is less obvious to me.  But I draw 
reassurance and encouragement from the fact that some schemes fail on 
their predictions for $m_{t}$ or $|V_{cb}|$.  The fact that failure is 
possible gives meaning to success.

To be sure, there are other plausible approaches to the origin of 
fermion masses, including composite models, schemes that assign a 
special role to the top quark, other mechanisms that communicate mass 
radiatively to the quarks and leptons, and---within the framework of 
string theory---the idea that the pattern of fermion masses is 
determined by the topology of extradimensional space.  But the general 
lesson I would draw from the exercise we have just undertaken is that 
a baffling pattern of masses observed at low energies may well arise 
from a simple and comprehensible pattern at high scales.  For a 
sampler of recent work along these lines, see 
\cite{Anderson:1993ba,Barbieri:1998qs,Albright:1999jv,Blazek:1999ue}

\section{Concluding Remarks}
Understanding the mechanism of electroweak symmetry breaking is the 
urgent challenge for particle physics over the next decade.  Searches 
for the Higgs boson and for electroweak physics beyond the standard 
model give form to the experimental programs at LEP~2, the Tevatron 
Collider, and the Large Hadron Collider, and guide the explorations 
of new accelerator projects.  Many possibilities for decisive 
observations lie before us.  Looking beyond the exploration of the 
1-TeV scale we see the problem of fermion mass, which is intimately 
linked to electroweak symmetry breaking, and may also lead us to 
higher scales.  I am very optimistic about the prospects for both 
theoretical and experimental progress.  I believe that within the 
next decade we will complete the gauge-theory revolution, and I 
confidently expect that nature will give us clues that will lead us 
to still greater understanding.

\section*{Acknowledgments}
It is my pleasure to thank Paco del Aguila and Fernando Cornet for 
organizing a stimulating and enjoyable week in Sierra Nevada.
Fermilab is operated by Universities Research 
Association Inc.\ under Contract No.\ DE-AC02-76CH03000 with the 
United States Department of Energy.


\begin{thebibliography}{99}
\bibitem{EQT} In the high-energy limit, an amplitude for longitudinal 
gauge-boson interactions may be replaced by a corresponding amplitude 
for the scattering of massless Goldstone bosons: $\mathcal{M}(W_{L},Z_{L}) = 
\mathcal{M}(w,z) + \mathcal{O}(M_{W}/\sqrt{s})$ The 
equivalence theorem can be traced to the work of John M.\  Cornwall, 
David N.\  Levin, and George Tiktopoulos, \pr{10}{1145}{74}, 
\ib{11}{972E}{75}.  It was applied to this problem by Lee, Quigg, and 
Thacker, Ref.\ \cite{lqt}, and developed extensively by Chanowitz and 
Gaillard, Ref. \cite{LT8}, and others.
\bibitem{vdBij} J. van der Bij, ``Can gravity play a role at the electroweak 
scale?'' \hepph{9507389}.
\end{thebibliography}
\end{document}